\documentclass[]{JFM-FLM_Au}
\usepackage{overpic}
\usepackage{hyperref}
\hypersetup{linkcolor=blue}
\usepackage[normalem]{ulem}

\newcommand{\po}{\boldsymbol{\omega}}
\newcommand{\pgamma}{\boldsymbol{\gamma}}
\newcommand{\ptau}{\boldsymbol{\tau}}
\newcommand{\ppsi}{\boldsymbol{\psi}}
\newcommand{\pPsi}{\boldsymbol{\Psi}}
\newcommand{\pphi}{\boldsymbol{\phi}}
\newcommand{\pO}{\boldsymbol{\Omega}}

\newcommand{\pa}{\boldsymbol{a}}
\newcommand{\pb}{\boldsymbol{b}}
\newcommand{\pe}{\boldsymbol{e}}
\newcommand{\pf}{\boldsymbol{f}}
\newcommand{\pgg}{\boldsymbol{g}}

\newcommand{\ppl}{\boldsymbol{l}}
\newcommand{\ppm}{\boldsymbol{m}}
\newcommand{\pn}{\boldsymbol{n}}
\newcommand{\pu}{\boldsymbol{u}}
\newcommand{\pw}{\boldsymbol{w}}
\newcommand{\px}{\boldsymbol{x}}
\newcommand{\pzero}{\boldsymbol{0}}
\newcommand{\pF}{\boldsymbol{F}}
\newcommand{\pM}{\boldsymbol{M}}

\newcommand{\pD}{\mathsfbi{D}}
\newcommand{\pT}{\mathsfbi{T}}
\newcommand{\pW}{\mathsfbi{W}}

\newcommand{\rd}{\textrm{d}}

\graphicspath{{figures/}}

\lefttitle{A.-K. Gao, C. Xie and X.-Y. Lu}
\righttitle{Weighted integral methods}

\title{Weighted integral methods for fluid force diagnostics in incompressible flows}

\author{An-Kang Gao\aff{1}, Chenyue Xie\aff{1} \and Xi-Yun Lu\aff{1}}

\affiliation{\aff{1}Department of Modern Mechanics, University of Science and Technology of China, Hefei 230026, Anhui, China}

\corresau{An-Kang Gao, \email{ankanggao@ustc.edu.cn}; Xi-Yun Lu, \email{xlu@ustc.edu.cn}}


\begin{document}
\maketitle

\begin{abstract}
Whilst surface stress integration remains the standard approach for fluid force evaluation, control-volume integral methods provide deeper physical insights through functional relationships between the flow field and the resultant force. In this work, by introducing a second-order tensor weight function into the Navier-Stokes equations, we develop a novel weighted-integral framework that offers greater flexibility and enhanced capability for fluid force diagnostics in incompressible flows. Firstly, in addition to the total force and moment, the weighted integral methods establish, for the first time, rigorous quantitative connections between the surface-stress distribution and the flow field, providing potential advantages for flexible body analyses. Secondly, the weighted integral methods offer alternative perspectives on force mechanisms, through vorticity dynamics or pressure view, when the weight function is set as divergence-free or curl-free, respectively. Thirdly, the derivative moment transformation (DMT)-based integral methods (Wu et al., J., Fluid Mech. vol. 576,  2007, 265–286) are generalised to weighted formulations, by which the interconnections among the three DMT methods are clarified. In the canonical problem of uniform flow past a circular cylinder, weighted integral methods demonstrate advantages in yielding new force expressions, improving numerical accuracy over original DMT methods, and enhancing surface-stress analysis. Finally, a force expression is derived that relies solely on velocity and acceleration at discrete points, without spatial derivatives, offering significant value for experimental force estimation. This weighted integral framework holds significant promise for flow diagnostics in fundamentals and applications.
\end{abstract}

\begin{keywords}
Authors should not enter keywords on the manuscript, as these must be chosen by the author during the online submission process and will then be added during the typesetting process (see \href{https://www.cambridge.org/core/journals/journal-of-fluid-mechanics/information/list-of-keywords}{Keyword PDF} for the full list).  Other classifications will be added at the same time.
\end{keywords}


\section{Introduction}
\label{sec:intro}
The force and moment exerted by the fluid are the core concern in the design and operation of aerial and underwater vehicles \citep{anderson2010aero,shyy2010recent,Bandyopadhyay2016}. According to the Br\'eguet range equation \citep{breguet1923calcul}, $$\text{Range} = \frac{U_\infty}{g}\frac{1}{TSFC} \frac{L}{D}\ln\left(\frac{\text{Initial aircraft weight}}{\text{Final aircraft weight}}\right)$$ where $U_\infty$ is the cruising speed, $g$ is the gravitational acceleration, and $TSFC$ (thrust-specific fuel consumption) is the fuel mass consumption rate per unit of thrust, the cruising range of an aircraft linearly depends on its lift-to-drag ratio $L/D$, motivating the pursuit of a high lift-to-drag ratio in modern commercial aircraft design. The moment, on the other hand, not only affects the stability of the vehicles but also serves as a primary parameter for the design of wing structures. In aquatic environments, the unsteady hydrodynamic force and moment present a major challenge for the control of autonomous underwater vehicles \citep{SAHOO2019145}. These examples demonstrate the enduring importance of accurate prediction and optimisation of fluid force and moment across engineering areas, from aerodynamic efficiency to underwater mobility.

By definition, the total force and moment experienced by a body submerged in the fluid are the integral of fluid stress $-p\hat{\pn}+\hat{\ptau}$ and its moment over the body surface $\partial B$,
\begin{equation}\label{eq:standardF}
    \pF  \coloneq \oint_{\partial B}{-p\hat{\pn} \rd S}  + \oint_{\partial B}{\hat{\ptau} \rd S},
\end{equation}
\begin{equation}\label{eq:standardM}
    \pM  \coloneq \oint_{\partial B}{-p\px\times\hat{\pn} \rd S} + \oint_{\partial B}{\px\times\hat{\ptau} \rd S}.
\end{equation}
Here, $p$ is the static pressure of the fluid, $\hat{\ptau}$ is the viscous stress, $\px$ is the spatial position, and $\hat{\pn}$ is the unit normal vector of $\partial B$ pointing from the solid side to the fluid side. The relevant definitions are schematically shown in figure~\ref{fig:controlvolume} and summarised in Appendix~\ref{app:nome}. Equations~\eqref{eq:standardF} and \eqref{eq:standardM} also provide a natural decomposition of force and moment into the pressure component and the viscous friction component, which are denoted by $\pF_p, \pM_p$ and $\pF_f, \pM_f$, respectively.

When the governing equations of the fluid dynamics can be simplified to permit analytical solutions, the fluid force and moment can be directly derived from the body geometry and kinematic parameters, establishing a direct relationship between force or moment and flow parameters. These scenarios typically apply to high-Reynolds-number flows without boundary layer separations, allowing the Euler equation to describe the external fluid motion, while the boundary layer theory \citep{prandtl1904uber} accounts for viscous effects near the solid wall. Using the potential flow theory, \citet{kutta1902lift} and \citet{joukowski1906chute} found that the lift force of a two-dimensional (2D) airfoil equals $L=\rho U_\infty\Gamma$ with $\Gamma$ being the circulation around the airfoil. \citet{munk1919isoperimetrische} developed a thin-airfoil theory, enabling direct calculation of lift and pitching moment from the airfoil geometry. For the finite-span wing in three-dimensional (3D) flows, by simplifying the wing as a series of quasi-2D airfoils of varying circulation, \citet{prandtl1918tragfl} developed a lifting line theory which predicts the lift and induced drag with good accuracy. 
\citet{theodorsen1935general} and \citet{von1938airfoil} considered the aeroelastic flutter of the thin airfoil and took into account the unsteady vortex shedding from the trailing edge. They developed a theory that predicts the unsteady force and moment of the airfoil undergoing small-amplitude oscillations. 

\begin{figure}
	\centerline{\includegraphics[width=0.8\textwidth]{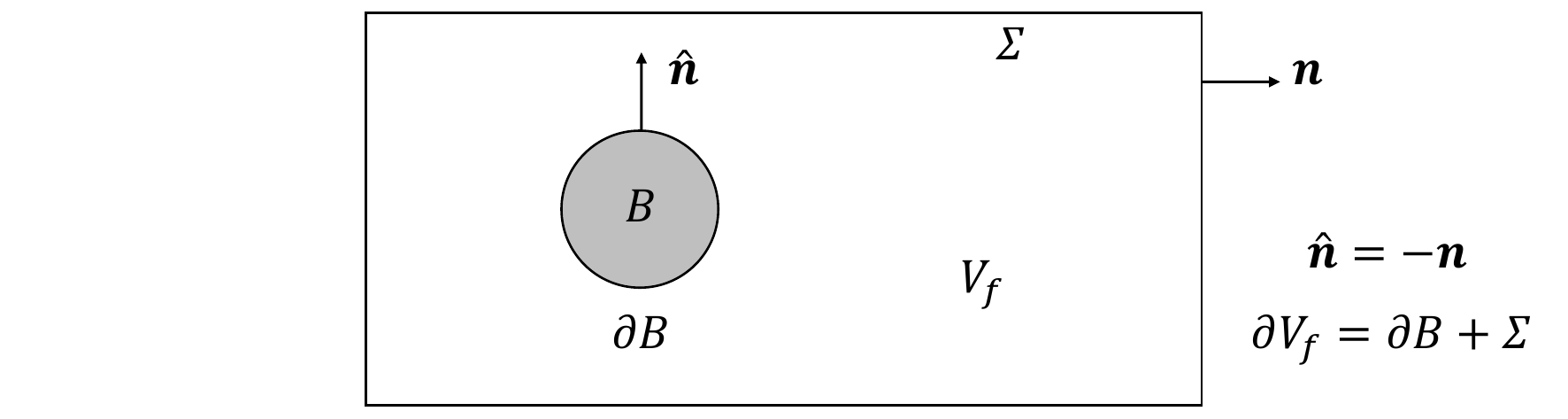}}
	\caption{Schematic diagram of the control volume enclosing the solid body.}
	\label{fig:controlvolume}
\end{figure}

In real viscous flows, the inherent nonlinearity of the Navier-Stokes (N-S) equations gives rise to diverse phenomena such as boundary layer separation from smooth surfaces, roll-up of free shear layers, self- and mutual-induced vortex motions, and vortex breakdown, precluding the possibility of analytical solutions. The coupling between fluid motion and the dynamics of solid structures further introduces nonlinearity on the boundaries. In most practical scenarios, it relies on computational fluid dynamics (CFD) and experimental measurements to obtain force, moment and flow field data. With advanced numerical methods, modern computational facilities, and experimental instruments such as Particle Image Velocimetry (PIV), massive quantities of data are generated. However, identifying dominating flow parameters, coherent structures, and key flow processes, as well as designing effective flow control methods, remains challenging \citep{wu2018fundamental}.

To diagnose the complex behaviour and multiple origins of fluid force and moment in viscous flows, control-volume integral methods that express force and moment as integrals over a control volume and its boundaries are developed. Comprehensive reviews of these methods and their applications can be found in \citet{wu2018fundamental}, \citet{graham2024vortex} and \citet{liu2024anual}. Except for the impulse theory \citep{burgers1920resistance,Lighthill1986theo,wu1981theory}, which has a time derivative outside the volume-integral term, these integral methods have a unified mathematical form as follows:
\begin{equation}\label{eq:intF}
    \pF  = \oint_{\partial B}{\pf_{\partial B} \rd S}  + \oint_{\Sigma}{\pf_\Sigma \rd S} + \int_{V_f}{\pf_{\text{vol}}\rd V},
\end{equation}
\begin{equation}\label{eq:intM}
    \pM  = \oint_{\partial B}{\ppm_{\partial B} \rd S}  + \oint_{\Sigma}{\ppm_\Sigma \rd S} + \int_{V_f}{\ppm_{\text{vol}}\rd V}.
\end{equation}
Here, $V_f$ is the control volume enclosing the solid body and $\Sigma$ is the outer boundary of $V_f$, as shown in figure~\ref{fig:controlvolume}. $\pf$ and $\ppm$ with different subscripts represent the corresponding integrands.

\begin{figure}
	\centerline{\includegraphics[width=1\textwidth]{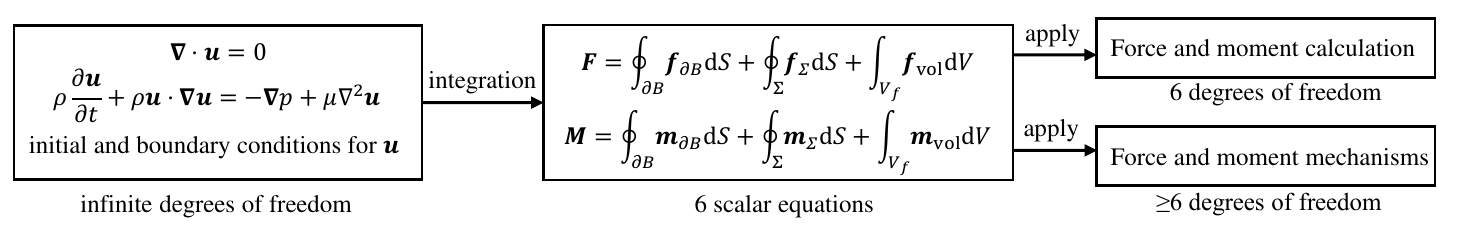}}
	\caption{Summary of the roles of control-volume integral methods for force and moment diagnostics.}
	\label{fig:explainwhy}
\end{figure}

The control-volume integral methods serve two critical roles in flow analysis, as summarised in figure~\ref{fig:explainwhy}. On the one hand, they provide complementary approaches to the standard surface-stress integral for force and moment calculations. For instance, the variational method developed by \citet{quartapelle1983force} eliminates explicit pressure dependence, enabling direct estimation of force from the PIV-measured velocity field \citep{diaz2022application}. On the other hand, these methods establish quantitative relationships between force and the flow field, offering interpretations of force generation mechanisms from the view of flow structures. A notable example is the elucidation of the drag-to-thrust transition mechanism in flapping airfoils through the vortex street transition from the B\'ernad-von K\'arm\'an vortex street to the reverse B\'ernad-von K\'arm\'an vortex street \citep{andersen2017wake}, achieved by applying the impulse theory. Nevertheless, current methodologies exhibit limitations in the following two aspects.

First, current control-volume integral methods provide exact formulas for the total force and moment but do not yield local stress distributions. Although total force and moment suffice to determine rigid body motion, local stress distributions are essential for the structural design and aeroelasticity analysis, especially in the case of the high altitude long endurance (HALE) aircraft with high-aspect-ratio flexible wings \citep{PalaciosCesnik2023}. Additionally, as demonstrated by \citet{protas2007attempt} and \citet{diaz2022application}, the spatial derivative of the fluid velocity near the body surface is often encountered in force expressions, posing significant challenges for experimental measurements due to the low resolution and high noise in near-wall regions. 
These limitations necessitate a systematic development of novel expressions for force and stress.

Second, when applied to the analysis of force and moment generation mechanisms, the diagnostic capability of an individual method is restricted by the limited information it captures. As indicated by figure~\ref{fig:explainwhy}, loss of information is inherent when deriving the six scalar equations for $\pF$ and $\pM$ from the governing equations via control volume integration. However, force mechanism analysis requires identifying contributions from local flow structures, distinct flow processes \citep{liu2014longitudinal}, or body kinematics, necessitating more degrees of freedom than the number of integral equations. Therefore, a systematic generalisation of integral methods, incorporating additional degrees of freedom, is needed to enhance their diagnostic capability. Furthermore, it is beneficial to combine complementary methods for a comprehensive understanding of flow mechanisms, for which a prerequisite is clarifying the specific information captured by each method and the interrelationships among them.

In this work, the concept of weighted integration, which arises from the finite element method for solving partial differential equations and also from the virtual power principle \citep{Yu2014virtual}, is employed to generalise the control-volume integral methods into weighted integral formulations. If the weight function is an arbitrary second-order tensor function, the weighted integration of the N-S equations yields the weak-form momentum balance equations. 
This weak form theoretically retains all information about fluid momentum transport and establishes functional mappings between the flow field and the surface stress distribution. Therefore, the weighted formulations provide a powerful framework for analysing the information captured by various integral methods, such as projection methods \citep{quartapelle1983force} and derivative moment transformation (DMT) methods \citep{wu2006vorticity,wu2007integral}, and elucidating their interrelationships. Furthermore, with additional degrees of freedom in the weight function, the weighted integral approach facilitates the systematic derivation of novel methods, enhancing flexibility and accuracy in force and moment estimations.

The remaining paper is organised as follows. In \S\ref{sec:review}, we give a brief review of control-volume integral methods for force and moment analysis. The weighted integral framework, including the general weighted methods, the weighted projection methods and the weighted DMT methods, is presented in \S\ref{sec:weight}. Numerical tests that support the validity and advantages of the weighted integral methods are provided in \S\ref{sec:numer}. Finally, in \S\ref{sec:conclusion}, we summarise the main conclusions.

\section{Review of control-volume integral methods}
\label{sec:review}
Force and moment diagnostics methods based on the momentum balance in a control volume have been extensively studied since the pioneering works of \citet{prandtl1918tragfl}, \citet{burgers1920resistance} and \citet{quartapelle1983force}. Comprehensive reviews of these methods are available in \citet{wu2018fundamental}, \citet{graham2024vortex} and \citet{liu2024anual}. Here, we briefly summarise methods of direct relevance to the present work, focusing on their concepts and underlying physics.

\subsection{Steady flow}
The vortex-force theory developed by \citet{prandtl1918tragfl} and refined by \citet{taylor1926note} and \citet{wu2006vorticity} represents a significant advancement in steady aerodynamics. In inviscid flows, \citet{prandtl1918tragfl} derived the aerodynamic force as
\begin{equation}\label{eq:vortexforce}
    \pF = \int_{V_{f\infty}} \rho\pu\times\po \rd V.
\end{equation}
Here, $\rho$ is the constant fluid density, $\pu$ is the velocity vector, $\po\coloneq\bnabla\times\pu$ is the vorticity vector, and $V_{f\infty}$ is the entire fluid domain. The volume integrand $\rho\pu\times\po$, which is perpendicular both to the local velocity and to the vorticity, is termed vortex force \citep{von1935general,saffman1992vortex} and equation~\eqref{eq:vortexforce} is referred to as the vortex-force theory \citep{alekseenko2007theory}. In 2D flows, equation~\eqref{eq:vortexforce} yields the Kutta-Joukowski theorem \citep{kutta1902lift,joukowski1906chute}, without drag component. In 3D flows, the streamwise component of equation~\eqref{eq:vortexforce} is generally non-zero, and it is identified as the induced drag, a drag component caused by the wake-vortex-induced downwash on the wing \citep{prandtl1918tragfl}. In viscous flows, \citet{taylor1926note} and \citet{wu2006vorticity} generalised the vortex-force theory by employing a finite control volume $V_f$ whose downstream surface is perpendicular to the incoming flow, enabling the identification of the induced drag and other force components. Applications of viscous steady vortex-force theory can be found in the work of \citet{Marongiu2010Far} and \citet{zou2019induced}.

\subsection{Unsteady flow}
Unsteady flows prevail in nature and engineering. They exist not only at high Reynolds number flows where the flow is turbulent, containing rich vortex dynamics, but also in flapping wings and unsteady incoming flows such as gusts. Force diagnostics in unsteady flows attracts great research interest due to its relevance to animal locomotion and bio-inspired design. It is also the main focus of the present study.

According to Newton's laws of motion, the interactive force exerted on the fluid from the solid body can be related to the time rate of change of the fluid momentum. However, the volume integral of momentum over the entire fluid domain does not converge in incompressible flows \citep{saffman1992vortex}, preventing the direct application of Newton's second law in the fluid system. This difficulty is overcome by \citet{burgers1920resistance}, \citet{wu1981theory} and \citet{Lighthill1986theo} independently through replacing the fluid momentum with the vorticity moment, which is also named vortical impulse. Using the DMT identities shown in Appendix \ref{app_DMT}, the fluid momentum in a finite control volume $V_f$ equals
\begin{equation}
    \int_{V_f}\rho\pu\rd V = \frac{1}{k}\int_{V_f}\rho\px\times\po\rd V - \frac{1}{k}\oint_{\partial {V_f}}\rho\px\times(\pn\times\pu)\rd S.
\end{equation}
Here, $k+1$ equals the space dimension and $\partial V_f$ is the boundary of the control volume $V_f$, as indicated in figure~\ref{fig:controlvolume}. If there is no vorticity outside the control volume $V_f$, it is proved that the force and moment experienced by the solid body are
\begin{eqnarray}
    \label{eq:ImpulseF}\pF &=& -\frac{\rho}{k}\frac{\rd}{\rd t}\int_{V_f}\px\times\po\rd V - \frac{\rho}{k}\frac{\rd}{\rd t}\oint_{\partial B}\px\times(\hat{\pn}\times\pu)\rd S, \\
    \label{eq:ImpulseM}\pM &=& \frac{\rho}{2}\frac{\rd}{\rd t}\int_{V_f}|\px|^2\po\rd V + \frac{\rho}{2}\frac{\rd}{\rd t}\oint_{\partial B}|\px|^2(\hat{\pn}\times\pu)\rd S.
\end{eqnarray}
Equation~\eqref{eq:ImpulseF} decomposes the force into the time derivative of the total vorticity moment and a surface-integral term that relates to the body's motion. Equation~\eqref{eq:ImpulseM} does a similar thing for the moment. They
are referred to as impulse theory or vorticity moment theory in the literature. The requirement that no vorticity exists outside the control volume $V_f$ is demanding in real applications, and therefore, variants of the impulse theory that apply to finite control volumes are developed \citep{noca1996evaluation,li2012force,Wang2013lift}. Notably, \citet{li2012force} took into account the contribution from vorticity outside $V_f$ and developed a finite-domain vorticity moment theory that applies to an arbitrary control volume. \citet{Kang2018mini} found that the control volume in the work of \citet{li2012force} can be minimised only to contain vorticity regions that connect the body, which greatly simplifies the force expression.

The impulse theory and its variants have achieved great success in applications, especially in elucidating the high-lift mechanism in insect flight. However, as noted by \citet{wu2006vorticity}, the time derivative in the impulse theory is outside the volume integral, obscuring a direct correlation between the force and flow structures. To address this problem, \citet{wu2007integral} derived three integral force formulas with time derivatives embedded inside the integrals: an advection form, a diffusion form, and a boundary form. These formulas allow for the visualisation of force contributions by analysing the spatial distributions of their integrands. Notably, the advection-form formula is also referred to as the unsteady vortex-force theory in \citet{wu2018fundamental} because it contains the volume integral of the vortex force $\rho \pu\times\po$, which resembles the inviscid vortex-force theory \citep{prandtl1918tragfl} given by equation~\eqref{eq:vortexforce}. While all three formulas accurately recover the total force in numerical test cases, their integrands exhibit remarkably distinct spatial distributions. Specifically, the vortex force is concentrated in the boundary layer, the free shear layer and the near-wake vortices. In contrast, the volume integrand of the diffusion-form formula is strongly localised near the solid wall.

By projecting the N-S equations onto a divergence- and curl-free vector function, \citet{quartapelle1983force} developed an alternative approach to associate the fluid force and moment with the velocity field in incompressible flows. First, a set of auxiliary harmonic functions $\eta_{\overline{i}}$ with $\overline{i}=1,2,3,4,5,6$ are introduced, which satisfy
$\nabla^2\eta_{\overline{i}} = 0$
in the fluid domain and boundary conditions
$\hat{\pn}\cdot\bnabla\eta_i = -\hat{\pn}\cdot\pe_i$ and
$\hat{\pn}\cdot\bnabla\eta_{i+3} = -(\px\times\hat{\pn})\cdot\pe_i$ with $i=1,2,3$
on the solid surface. At the infinitely far field, $\eta_{\overline{i}}$ goes to zero. The pressure force $\pF_p$ and the moment due to pressure $\pM_p$ can be expressed as
\begin{equation}\label{eq:intpphiF}
    \pF_p\cdot\pe_i \coloneq \oint_{\partial B}-p\hat{\pn}\cdot\pe_i\rd S = \oint_{\partial B} \hat{\pn}\cdot(p\bnabla\eta_i) \rd S = -\int_{V_{f\infty}} \bnabla p\cdot\bnabla\eta_i \rd V,
\end{equation}
\begin{equation}\label{eq:intpphiM}
    \pM_p\cdot\pe_i \coloneq \oint_{\partial B}-p(\px\times\hat{\pn})\cdot\pe_i\rd S = \oint_{\partial B} \hat{\pn}\cdot(p\bnabla\eta_{i+3}) \rd S = -\int_{V_{f\infty}} \bnabla p\cdot\bnabla\eta_{i+3} \rd V.
\end{equation}
Here, $\pe_i, i=1,2,3$ are the unit basis vectors of the coordinate system. The force and moment can be subsequently obtained by the volume integral of the dot product of the N-S equations and $\bnabla\eta_{\overline{i}}$. After applying the divergence theorem, the remaining volume integrand is $\rho(\pu\cdot\bnabla\pu)\cdot\bnabla\eta_{\overline{i}}$. \citet{protas2007attempt} further theoretically examined this method and found that the vorticity in the body-surface integral cannot be eliminated, which is problematic in experimental applications. It is noted that the gradient of the auxiliary function $\eta_i$ can be explained as a virtual velocity in the virtual power principle developed by \citet{Yu2014virtual}.

Following the work of \citet{quartapelle1983force}, a variety of projection-based force diagnostics methods have been developed. Recognising that the vorticity is more localised compared with the velocity gradient, \citet{howe_1989} and \citet{chang1992potential} reformulated the advection term as $\rho\pu\cdot\bnabla\pu = \rho\po\times\pu+\bnabla(\rho|\pu|^2/2)$ and derived the force and moment expressions with $(\rho\po\times\pu)\cdot\bnabla\eta_{\overline{i}}$ as the volume integrand. 
\citet{Yu2014virtual} and \citet{gao2019note} noted that the Galilean invariance, which is a fundamental requirement for the force in Newtonian mechanics, is not inherently guaranteed in most force diagnostics methods due to their velocity dependence. To address this limitation, \citet{gao2019passing} derived a force expression with a Galilean-invariant volume integrand $2\rho Q\eta_i$, which is obtained from the identity $\bnabla\cdot(\rho\pu\cdot\bnabla\pu) = -2\rho Q$ with $Q$ being the second invariant of the velocity gradient tensor. The moment expression based on $2\rho Q\eta_{i+3}$ is obtained by \citet{menon2021quantitative}. \citet{li2018vortex} and \citet{li2020vortex} utilised the force expression of \citet{howe_1989} and developed a vortex force and moment map method that links the vortex motion to the force and moment. These theoretical methods have found wide applications, to name a few, in numerical simulations, see \citet{martine2019assessment} and \citet{chiu2023vorticity}, and in experiments, see \citet{diaz2022application}, \citet{gehlert2023vortex} and \citet{otomo2025vortex}.

\section{The weighted integral methods for force diagnostics}
\label{sec:weight}
The externally unbounded incompressible flow past a solid body $B$ is considered. The flow is governed by the continuity equation and the N-S equations
\begin{equation}\label{eq:continous}
    \bnabla\cdot \pu = 0,
\end{equation}
\begin{equation}\label{eq:momentequation}
    \rho\pa\coloneq \rho\frac{\partial\pu}{\partial t}+\rho\pu\cdot\bnabla\pu = - \bnabla p +  \bnabla\cdot\pT.
\end{equation}
Here, $\pa$ is the fluid acceleration, $\pT\coloneq 2\mu\pD$ is the viscous stress tensor with $\mu$ being the dynamic viscosity, and $\pD\coloneq 0.5\left(\bnabla\pu+(\bnabla\pu)^T\right)$ is the symmetric strain rate tensor.
The velocity satisfies the no-slip boundary condition on the solid surface $\partial B$ and a proper far-field boundary condition. With a prescribed initial condition for $\pu$, the evolution of the incompressible flow can be determined.

The control volume $V_f$ enclosing the body is schematically shown in figure~\ref{fig:controlvolume}. The boundary of the control volume is denoted by $\partial V_f$ which is constituted by the body surface $\partial B$ and the outer boundary $\Sigma$. $\pn$ is the unit normal vector of $\partial V_f$ pointing outside the control volume and there is $\pn=-\hat{\pn}$. The viscous stress exerted on the solid body is denoted by $\hat{\ptau}\coloneq \hat{\pn}\cdot\pT$. Meanwhile, the viscous stress exerted on the fluid surface is $\ptau\coloneq\pn\cdot\pT$.

\subsection{General weighted method}
\label{sec:generalweight}
Assuming $\pW=W_{ij}\pe_i\pe_j$ is a smooth second-order tensor function defined in the fluid domain as well as on the body surface, the weighted momentum equation is derived as
\begin{equation}\label{eq:Weightmoment}
    \rho\pa\cdot\pW =  \bnabla\cdot(-p\pW+\pT\cdot\pW) + p\bnabla\cdot\pW - \pT:\bnabla\pW,
\end{equation}
with $\pT:\bnabla\pW=T_{ij}(\partial W_{ik}/\partial x_j)\pe_k$. 
The weighted force $\pF_w$ is defined by
\begin{equation}\label{eq:surfFw}
    \pF_w =  \pF_{w,p} + \pF_{w, f} \coloneq \oint_{\partial B}{-p\hat{\pn}\cdot\pW \rd S}  + \oint_{\partial B}{\hat{\ptau}\cdot\pW \rd S},
\end{equation}
in which $\pF_{w,p}$ and $\pF_{w, f}$ denote the weighted pressure force and the weighted friction force, respectively. 

The weight function $\pW$, interpreted as the dyad of virtual velocity in the virtual power principle \citep{Yu2014virtual}, also indicates the relative importance of local spatial regions. By increasing $W_{ij}$ on a local surface area while reducing its magnitude $|W_{ij}|$ in other regions, the local stress distribution is effectively represented through $\pF_w$. 
The total force and moment can also be recovered. If $\pW=\delta_{ij}\pe_i\pe_j$ on the body surface with $\delta_{ij}$ being the Kronecker delta, $\pF_w$ equals the total fluid force, $\pF$; whereas if $\pW=\epsilon_{ijk}x_i\pe_j\pe_k$ on $\partial B$ with $\epsilon_{ijk}$ being the permutation tensor, $\pF_w$ recovers the total moment, $\pM$. Furthermore, if the body is rigid and $\pW$ equals $\hat{\pn}\hat{\pn}$ on $\partial B$, $\pF_w$ becomes the pressure force, $\pF_p$, because $\hat{\ptau}$ is perpendicular to $\hat{\pn}$. On the contrary, if $\pW=\delta_{ij}\pe_i\pe_j-\hat{\pn}\hat{\pn}$ on $\partial B$, $\pF_w$ equals the friction force, $\pF_f$.

To establish quantitative relationships between the weighted force $\pF_w$ and the flow field, equation~\eqref{eq:Weightmoment} is integrated over the control volume $V_f$. After applying the divergence theorem to the term $\int_{V_f}\bnabla\cdot(-p\pW+\pT\cdot\pW)\rd V$, the general weighted method is obtained as follows,
\begin{equation}\label{eq:generalweight}
    \pF_w = \int_{V_f}{(-\rho\pa\cdot\pW + p\bnabla\cdot\pW - \pT:\bnabla\pW) \rd V} + \oint_{\Sigma}{(-p\pn\cdot\pW+\ptau\cdot\pW) \rd S}.
\end{equation}
The last term in the above equation is the weighted surface-stress integral over $\Sigma$ and it is denoted by $\pF_{w,\Sigma}$,
\begin{equation}\label{eq:outerstress}
   \pF_{w,\Sigma} \coloneq \oint_{\Sigma}{(-p\pn\cdot\pW+\ptau\cdot\pW) \rd S}.
\end{equation}
In addition, the volume integral of $-\pT:\bnabla\pW$ in equation~\eqref{eq:generalweight}, denoted by $\pF_{w,\text{vis}}$, arises from the viscous effect and it can be transformed into the following four forms by applying the divergence theorem,
\begin{subequations}\label{eq:fwvis}
\begin{align}
\label{eq:fwvis1} \pF_{w,\text{vis}} &\coloneq \int_{V_f}{ -\pT:\bnabla\pW \rd V} = -2\mu\int_{V_f}{ \pD:\bnabla\pW \rd V},\\
\label{eq:fwvis2} \pF_{w,\text{vis}} &=-\mu\int_{V_f}{ (\bnabla\times\po)\cdot\pW \rd V}- \oint_{\partial V_f}{\ptau\cdot\pW\rd S}, \\
\label{eq:fwvis3} \pF_{w,\text{vis}} &= -\mu\int_{V_f}{ \po\cdot(\bnabla\times\pW) \rd V}+ 2\mu \oint_{\partial V_f}{\left[ (\pn\cdot\pu)(\bnabla\cdot\pW) - \pn\cdot(\pu\cdot\bnabla\pW) \right]\rd S},\\
\label{eq:fwvis4} \pF_{w,\text{vis}} &= \mu\int_{V_f}{ \pu\cdot\left[2\bnabla(\bnabla\cdot\pW) - \bnabla\times(\bnabla\times\pW)\right] \rd V} - \mu \oint_{\partial V_f}{(\pu\pn+\pn\pu):\bnabla\pW\rd S}.
\end{align}
\end{subequations}
The mathematical details are provided in Appendix \ref{app:Fwvis}.
Equations~\eqref{eq:fwvis1}-\eqref{eq:fwvis4} involve spatial derivatives of different orders for the velocity and the weight function. A proper choice of the form of $\pF_{w,\text{vis}}$ offers convenience in applications. For instance, if $\pW$ has an analytical expression while $\pu$ is known only at discrete points, equation~\eqref{eq:fwvis4} can be utilised to avoid the numerical differentiation of $\pu$. It should be noted that the weighted friction force $\pF_{w,f}$ cancels out in equation~\eqref{eq:generalweight} when equation~\eqref{eq:fwvis2} is employed.

In the general weighted method, the weight function $\pW$ is set as a second-order tensor to maximise the flexibility of the formula and also to ensure terms in equation~\eqref{eq:generalweight} are vectors. If equation~\eqref{eq:generalweight} is projected onto a basis vector $\pe_j$, this is equivalent to having a vector weight function $\pw=W_{ij}\pe_i$. Additionally, when $\pW$ is an isotropic tensor, i.e., $W_{ij}=w\delta_{ij}$, this setup is equivalent to a scalar weight function $w$.

If the weight function $\pW$ is taken as an arbitrary function, the general weighted method is equivalent to the weak-form momentum balance equations, which contain all information of momentum transport in the control volume. However, one drawback of the general weighted method is that the pressure is involved in the volume integration. Because the pressure field is hard to measure in experiments and the pressure only serves as a Lagrangian multiplier that ensures the incompressible condition \citep{chorin1993mathematical}, it is preferred to eliminate pressure by restricting the weight function to a specific function space.

\subsection{Weighted projection methods}
\label{sec:weightproject}
The fluid motion can be decomposed into a transverse process, described by the vorticity transport equation, and a longitudinal process, characterised by pressure wave propagation, as established by \citet{chu1958non}, \citet{liu2014longitudinal} and \citet{mao2020study}. In incompressible flows, the longitudinal process is governed by the pressure Poisson equation owing to the infinite speed of sound. In this section, we demonstrate that the general weighted method~\eqref{eq:generalweight} captures the transverse process when $\pW$ is divergence-free and the longitudinal process when $\pW$ is curl-free. 

\subsubsection{Divergence-free weight function}
Following the methodology of \citet{quartapelle1983force}, where the N-S equations are projected onto a divergence-free vector field, we assume the weight function satisfies $\bnabla\cdot\pW=\pzero$. Consequently, the term $p\bnabla\cdot\pW$ vanishes in equation~\eqref{eq:generalweight}. 
The weighted projection method with a divergence-free weight function is given by
\begin{equation}\label{eq:weightedproject}
    \pF_w = \int_{V_{f}}{-\rho\pa\cdot\pW \rd V}  +\pF_{w,\text{vis}} + \pF_{w,\Sigma}.
\end{equation}
As proposed by \citet{howe_1989} and \citet{chang1992potential}, replacing the acceleration $\pa$ with $\partial\pu/\partial t + \po\times\pu+\bnabla(0.5|\pu|^2)$ yields a localised integrand for uniform incoming flows, and hence an alternative form of weighted projection method with a divergence-free weight function is derived as
\begin{equation}\label{eq:weightedprojectlamb}
    \pF_w = \int_{V_{f}}{\left[-\rho\frac{\partial \pu}{\partial t}\cdot\pW+\rho(\pu\times\po)\cdot\pW\right] \rd V} - \oint_{\partial V_f}{0.5\rho|\pu|^2\pn\cdot\pW\rd S}  +\pF_{w,\text{vis}} + \pF_{w,\Sigma}.
\end{equation}
In equation~\eqref{eq:weightedprojectlamb}, the weighted vortex force term $\rho(\pu\times\po)\cdot\pW$ in the volume integrand underscores the contribution of the vortex force to the generation of aerodynamic or hydrodynamic force.

If we consider a subspace of the divergence-free weight functions whose total flux over the body surface is zero,
\begin{equation}
    \oint_{\partial B}{\hat{\pn}\cdot\pW\rd S} = \pzero,
\end{equation}
by vector calculus, this subspace is constituted by $\bnabla\times\pPsi$ with $\pPsi$ being an arbitrary second-order tensor function. As demonstrated in Appendix \ref{app:vortransport}, equation~\eqref{eq:weightedproject} can be cast into the following equation
\begin{equation}
\begin{split}
    \int_{V_f}{\left(\rho\frac{\partial\po}{\partial t}+\rho\pu\cdot\bnabla\po-\rho\po\cdot\bnabla\pu - \mu\nabla^2\po\right)\cdot\pPsi\rd V} + & \\
    \oint_{\partial V_f}{\left[\mu\pn\cdot\bnabla\po -\left( \mu(\pn\times\bnabla)\times\po +\rho\pn\times\pa+\pn\times\bnabla p\right)\right]\cdot\pPsi\rd S}&=0,
\end{split}
\end{equation}
in which the volume integral reveals the vorticity transport equation and the surface integral yields the boundary vorticity flux (BVF) that measures the vorticity creation rate from the boundary \citep{Wu_Wu_1993}. Therefore, the weighted projection method with a divergence-free weight function, equation~\eqref{eq:weightedproject} or \eqref{eq:weightedprojectlamb}, captures all the information of vorticity evolution and vorticity generation on the wall, i.e., the transverse process.

\subsubsection{Curl-free weight function}
If the weight function is curl-free and it has no circulation for multi-connected fluid domains, the weight function can be expressed as the gradient of a vector potential, $\pW=\bnabla\pphi$. Here, $\pphi=\phi_i\pe_i$. Using vector identities, $-\pa\cdot\bnabla\pphi=-\bnabla\cdot(\pa\pphi)+(\bnabla\cdot\pa)\pphi$ and $-(\bnabla\times\po)\cdot\bnabla\pphi=\bnabla\cdot\left[(\nabla^2\pu)\pphi\right]$, and noting $\bnabla\cdot\pa=-2Q$, where $Q$ is the second invariant of the velocity gradient tensor, the weighted pressure force $\pF_{w,p}$ can be obtained from equation~\eqref{eq:generalweight} by employing equation~\eqref{eq:fwvis2} for $\pF_{w,\text{vis}}$ as
\begin{equation}\label{eq:potentialprojectvector}
\begin{split}
\pF_{w,p} = & \int_{V_{f}}{\left(-2\rho Q \pphi + p\nabla^2\pphi\right) \rd V} + \oint_{\partial V_f}{\pn\cdot(-\rho\pa+\mu\nabla^2\pu)\pphi \rd S} \\
&+ \oint_{\Sigma}{-p\pn\cdot\bnabla\pphi \rd S}.
\end{split}
\end{equation}
Equation~\eqref{eq:potentialprojectvector} is the weighted projection method with a curl-free weight function. Note that the pressure is not eliminated in the volume integral because equation~\eqref{eq:potentialprojectvector} establishes a quantitative relationship between $2\rho Q$, which is the source term of the pressure Poisson equation, and the pressure throughout the control volume and on the boundary $\partial V_f$.

As illustrated in Appendix~\ref{app:pressurepoisson}, equation~\eqref{eq:potentialprojectvector} can be transformed into
\begin{equation}
\int_{V_{f}}{\phi_i(\nabla^2p-2\rho Q) \rd V}+\oint_{\partial V_f}{\phi_i\pn\cdot(-\bnabla p - \rho\pa+\mu\nabla^2\pu)\rd S} = 0,
\end{equation}
where the volume integral yields the pressure Poisson equation and the surface integral reveals the Neumann pressure boundary condition. Therefore, equation~\eqref{eq:potentialprojectvector} is equivalent to the pressure Poisson equation together with the pressure boundary condition, i.e. the longitudinal process.

If the weight function is both divergence- and curl-free, equation~\eqref{eq:weightedproject} and equation~\eqref{eq:potentialprojectvector}, representing the transverse process and the longitudinal process, respectively, are both applicable. In this case, the pressure source term $2\rho Q$ in equation~\eqref{eq:potentialprojectvector} remains as the sole volume integrand, revealing the intrinsic relationship that the transverse process serves as the source of the longitudinal process.

\subsection{Weighted DMT methods}
\label{sec:weightDMT}
In this section, the three DMT methods \citep{wu2007integral} are generalised to weighted formulations to enhance their capability and elucidate their interconnections.
\subsubsection{Advection form}
First, three vectors $\pf_w$, $\ppl_w$ and $\pb_w$ are introduced, which are
\begin{equation}\label{eq:deffw}
    \rho \pf_w \coloneq \rho \pa\cdot\pW   - p\bnabla\cdot\pW + \pT:\bnabla \pW,
\end{equation}
the weighted Lamb vector $\ppl_w\coloneq(\po\times\pu)\cdot\pW$, and
\begin{equation}\label{eq:defbw}
    \rho \pb_w \coloneq \rho\pf_w-\rho\ppl_w = \rho \frac{\partial \pu}{\partial t}\cdot\pW +\bnabla( 0.5\rho|\pu|^2)\cdot\pW  - p\bnabla\cdot\pW + \pT:\bnabla \pW.
\end{equation}
The weighted N-S equations \eqref{eq:Weightmoment} can be rewritten as
\begin{equation}\label{eq:weightNSfw}
    \rho\pf_w = \bnabla\cdot(-p\pW+\pT\cdot\pW).
\end{equation}
It is noted that the volume integrand in the general weighted method \eqref{eq:generalweight} equals $-\rho\pf_w$ and that $\rho(\pb_w-\pf_w) = \rho(\pu\times\po)\cdot\pW$ is the weighted vortex force.

The advection-form weighted DMT method is obtained by applying DMT identity~\eqref{eq:DMTvector} to equation~\eqref{eq:generalweight},
\begin{equation}\label{eq:AdvectionDMT}
\begin{split}
 \pF_w = & \int_{V_f}{\left[-\frac{\rho}{k}\px\times(\bnabla\times\pb_w)+\rho(\pu\times\po)\cdot\pW\right] \rd V}  -\frac{\rho}{k}\oint_{\partial V_f}{\px\times(\pn\times\ppl_w) \rd S} \\
 &+ \pF_{w,\text{DMT},B} + \pF_{w,\text{DMT},\Sigma}.
\end{split}
\end{equation}
Here, the surface integrals are
\begin{equation}\label{eq:FB}
  \pF_{w,\text{DMT},B} \coloneq \frac{\rho}{k}\oint_{\partial B}{\px\times(\pn\times\pf_w) \rd S},
\end{equation}
\begin{equation}\label{eq:FSigma}
  \pF_{w,\text{DMT},\Sigma} \coloneq \frac{\rho}{k}\oint_{\Sigma}{\px\times(\pn\times\pf_w) \rd S} + \oint_\Sigma{(-p\pn\cdot\pW+\ptau\cdot\pW) \rd S}.
\end{equation}
In equation~\eqref{eq:AdvectionDMT}, the weighted vortex force contributes to the volume integral. Additionally, the pressure exists in the volume integrand in the form of $p\bnabla\cdot\pW$, which is generally non-zero for arbitrary weight functions.

Since no simplifications are made in the derivation of equation~\eqref{eq:AdvectionDMT} and the weight function $\pW$ is arbitrary, the advection-form weighted DMT method contains the same information as the general weighted method \eqref{eq:generalweight}, which represents the weak-form momentum balance equations.

\subsubsection{Diffusion and boundary forms}
An isotropic weight function is required to derive the diffusion-form and boundary-form DMT methods, i.e., $W_{ij}=w\delta_{ij}$. 

Following the work of \citet{wu2007integral}, the diffusion-form weighted DMT method is derived by substituting the weighted N-S equations~\eqref{eq:weightNSfw} into the volume and outer surface integrals of the advection-form equation~\eqref{eq:AdvectionDMT}. The mathematical details can be found in Appendix~\ref{app:diffu}. Here, the weighted force is expressed as
\begin{equation}\label{eq:diffusion}
 \pF_w = -\frac{1}{k}\int_{V_f}{\px\times\left[\bnabla\times(\bnabla\cdot(w\pT))\right] \rd V} + \pF_{w,\text{DMT},B} + \tilde{\pF}_{w,\text{DMT},\Sigma},
\end{equation}
with $\tilde{\pF}_{w,\text{DMT},\Sigma}$ being
\begin{equation}\label{eq:FSigma3}
  \tilde{\pF}_{w,\text{DMT},\Sigma} \coloneq \frac{1}{k}\oint_{\Sigma}{\px\times\left[\pn\times\left(\bnabla\cdot(w\pT)\right)\right] \rd S} + \oint_\Sigma{w\ptau \rd S}.
\end{equation}

In the derivation of equation~\eqref{eq:diffusion}, the weighted N-S equations are used for a second time in the control volume and on the outer surface, which leads to the cancellation of information in $V_f$ and on $\Sigma$. As demonstrated in Appendix~\ref{app:diffuinsensitive}, the diffusion-form method predicts the same force value even when the flow velocity and pressure are arbitrarily disturbed in the interior of the fluid domain, indicating that equation~\eqref{eq:diffusion} contains only the information on momentum balance on the body surface $\partial B$.

The boundary-form DMT method is obtained by substituting the weighted N-S equations~\eqref{eq:weightNSfw} into the surface integral of equation~\eqref{eq:diffusion}. The mathematical details can also be found in Appendix~\ref{app:diffu}. Here, the weighted force is
\begin{equation}\label{eq:boundaryform}
  \pF_w = \frac{1}{k}\oint_{\partial B}{\px\times[\hat{\pn}\times\bnabla(wp)] \rd S} + \oint_{\partial B}{w\hat{\ptau} \rd S}.
\end{equation}
Through DMT identity (\ref{eq:DMTscalar}), there is
\begin{equation}\label{eq:DMTidentity}
   \pF_{w,p}=\oint_{\partial B}{-w p\hat{\pn}  \rd S} = \frac{1}{k}\oint_{\partial B}{\px\times[\hat{\pn}\times\bnabla(w p)] \rd S}.
\end{equation}
Therefore, the boundary-form DMT method is equivalent to the weighted surface-stress integral~\eqref{eq:surfFw}, which reflects the fluid stress distribution on the $\partial B$.

If $B$ is a rigid body, the generalised Caswell formula \citep{wu2006vorticity} shows $2\pD=\hat{\pn}(\po_r\times\hat{\pn})+(\po_r\times\hat{\pn})\hat{\pn}$, in which $\po_r\coloneq\po-2\pO$ is the relative vorticity and $\pO$ is the angular velocity of the rigid body. Hence, the viscous stress equals $\hat{\ptau}=\mu\po_r\times\hat{\pn}$. Using the DMT identities~\eqref{eq:DMTcurl} and \eqref{eq:DMTscalar}, $\pF_{w,f}$ can be rewritten as
\begin{subequations}\label{eq:Ffdmt}
\begin{align}
\label{eq:ff1} \pF_{w,f} &= \mu\oint_{\partial B}{\px\times[(\hat{\pn}\times\bnabla)\times(w\po_r)] \rd S}, \text{ in 3D space},\\
\label{eq:ff2} \pF_{w,f} &= -\mu\oint_{\partial B}{\px\left[(\pe_z\times\hat{\pn})\cdot \bnabla(w\omega_{rz}) \right]\rd S}, \text{ in 2D space.}
\end{align}
\end{subequations}
In 2D space, $\po_r=\omega_{rz}\pe_z$.

\subsection{Discussions}
A summary of the weighted integral methods developed in this work is presented in Table~\ref{tab:summary}, along with the information captured by each method. First, the weighted surface-stress integral \eqref{eq:surfFw} generalises the standard surface-stress integral \eqref{eq:standardF} with additional degrees of freedom in the weight function, enabling evaluation of the fluid stress distribution over the body surface $\partial B$. Second, by restricting the weight function to be divergence-free or curl-free, the projection methods~\eqref{eq:weightedproject} and \eqref{eq:potentialprojectvector} capture the transverse process and the longitudinal process, governed by the vorticity transport equation and the pressure Poisson equation, respectively. Specifically, the divergence-free weight function eliminates the pressure term in the volume integral. Third, the three forms of weighted DMT methods are not equivalent. The advection form \eqref{eq:AdvectionDMT} contains the momentum balance equations, i.e. the N-S equations, in the control volume $V_f$ and on the boundary $\partial V_f$, whereas the diffusion form \eqref{eq:diffusion} captures only the momentum balance on the body surface $\partial B$. The boundary form \eqref{eq:boundaryform} is equivalent to the weighted surface-stress integral \eqref{eq:surfFw} that reveals the stress distribution on the body surface. Finally, time derivatives in the methods presented in this work are all embedded inside the integral. A weighted generalisation of the impulse theory, which has the time derivative outside the integral, remains unexplored here.

\begin{table}
  \begin{center}
\def~{\hphantom{0}}
  \begin{tabular}{p{3cm}|p{1.7cm}p{2.7cm}p{4.7cm}}
 Method & Equation & Weight function & Information captured\\
 \hline
 Weighted surface-stress integral & \eqref{eq:surfFw} & Arbitrary second-order tensor function & Fluid stress distribution on $\partial B$ \\
 \hline
 General weighted method & \eqref{eq:generalweight} & Arbitrary second-order tensor function & N-S equations in $V_f$ and momentum balance on $\partial V_f$\\
 \hline
 Weighted projection method with divergence-free weight function & \eqref{eq:weightedproject} \par or \par \eqref{eq:weightedprojectlamb} & $\bnabla\cdot\pW=\pzero$ & At least vorticity transport equation in $V_f$ and boundary vorticity flux on $\partial V_f$ \\
 \hline
 Weighted projection method with curl-free weight function& \eqref{eq:potentialprojectvector} & $\pW=\bnabla\pphi$, $\pphi$ is arbitrary vector function & Pressure Poisson equation in $V_f$ and Neumann pressure boundary condition on $\partial V_f$\\
 \hline
 Advection-form weighted DMT method & \eqref{eq:AdvectionDMT} & Arbitrary second-order tensor function & N-S equations in $V_f$ and momentum balance on $\partial V_f$ \\
 \hline
 Diffusion-form weighted DMT method & \eqref{eq:diffusion} & $W_{ij}=w\delta_{ij}$, $w$ is arbitrary scalar function & Momentum balance on $\partial B$ \\
 \hline
 Boundary-form weighted DMT method & \eqref{eq:boundaryform} & $W_{ij}=w\delta_{ij}$, $w$ is arbitrary scalar function & Fluid stress distribution on $\partial B$ \\
 \hline
  \end{tabular}
  \caption{Summary of the weighted integral methods.}
  \label{tab:summary}
  \end{center}
\end{table}

The weighted integral methods recover the original control-volume integral methods when special weight functions are adopted. If $W_{ij}=-\partial \eta_j/\partial x_i$, the weighted projection method with divergence-free weight function \eqref{eq:weightedprojectlamb} reduces to the force expression developed by \citet{chang1992potential}, and the weighted projection method with curl-free weight function \eqref{eq:potentialprojectvector} becomes the force expression developed by \citet{gao2019passing}. The case is also valid for the moment when $\eta_{j+3}$ is employed. Here, $\eta_j$ and $\eta_{j+3}$ that appear in equations~\eqref{eq:intpphiF} and \eqref{eq:intpphiM} are first introduced by \citet{quartapelle1983force}.
For the weighted DMT methods, if $\pW$ is a second-order unit tensor, i.e., $W_{ij}=\delta_{ij}$, extra terms involving spatial derivatives of $\pW$ vanish and, as a result, the original DMT methods are recovered.


It is widely acknowledged that the force in Newtonian mechanics is independent of the inertial frame of reference, known as Galilean invariance. The Galilean invariance in fluid force decomposition has been addressed by \citet{Yu2014virtual}, who claimed that each integral term in the force expression needs to be Galilean invariant, and by \citet{gao2019note}, who emphasised the Galilean invariance of integrals over local fluid regions. In aerodynamic and hydrodynamic force diagnostics, the Galilean invariance can be satisfied at three levels. First, for a prescribed weight function, the total weighted force $\pF_w$ is Galilean invariant, which is already guaranteed owing to the rigorous derivation of all methods. Second, when evaluating force contributions from local flow structures, the local integrals must be independent of the inertial frame of reference. Third, when analysing the spatial distributions of volume or surface integrands, the relevant integrands must be Galilean invariant. For example, integrands in equations~\eqref{eq:surfFw}, \eqref{eq:generalweight}, \eqref{eq:weightedproject} and \eqref{eq:potentialprojectvector} are all Galilean invariant, provided that $\pW$ is Galilean invariant. The second or third level is not necessarily satisfied in a specific application when only the total force is concerned. A practical criterion is that the focused force or force contribution should be Galilean invariant.

\section{Numerical tests}
\label{sec:numer}
\subsection{2-D flow past a circular cylinder}
The uniform flow past a stationary circular cylinder is employed to assess the performance of various weighted integral methods. The incoming flow velocity $U_\infty$, the fluid density $\rho$ and the cylinder diameter $d$ are designated as reference quantities. The Reynolds number is defined as $Re=\rho U_\infty d/\mu=100$. The fluid velocity, expressed as $\pu=u_x\pe_x+u_y\pe_y$, equals $U_\infty\pe_x$ at the infinitely far field and zero on the cylinder surface. The pressure at infinity, set to $p_\infty = 0$, is adopted as the reference pressure.
\begin{figure}
\centerline{\includegraphics[width=1\textwidth]{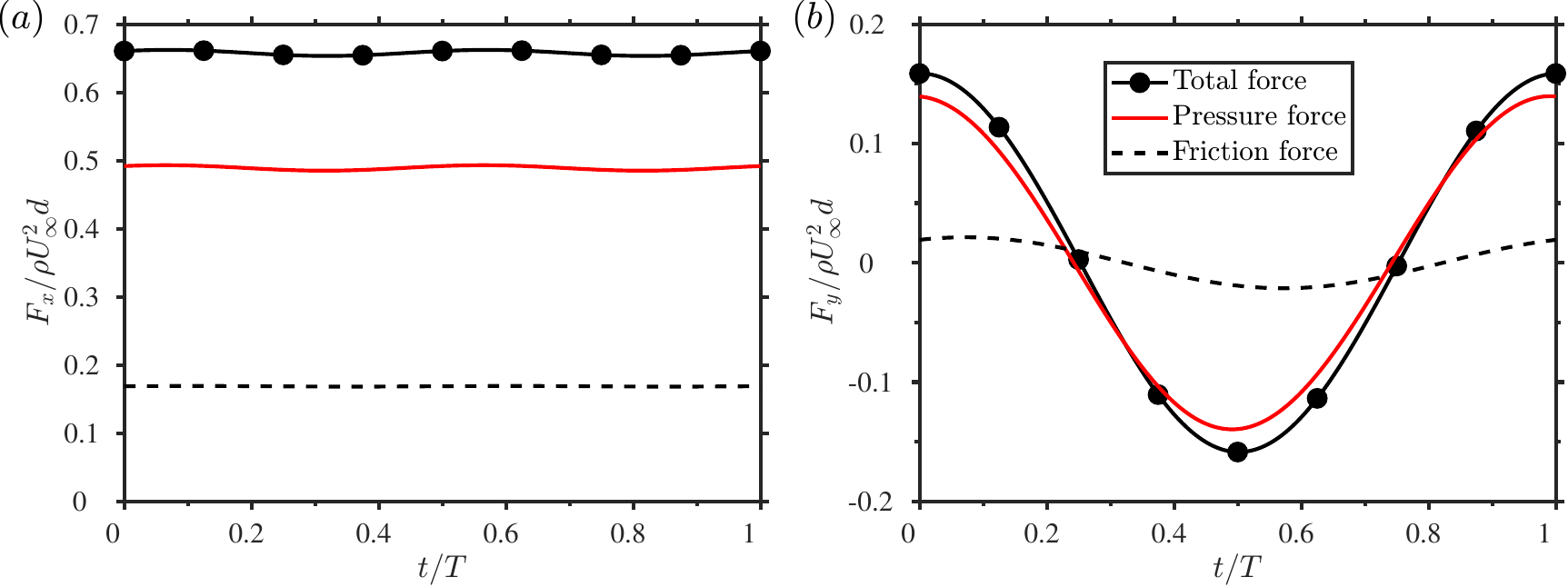}}
	\caption{Time-dependent dimensionless force on the circular cylinder at $Re=100$. (a) Drag. (b) Lift.}
	\label{fig:timeforce}
\end{figure}

\begin{figure}
\centerline{\includegraphics[width=1\textwidth]{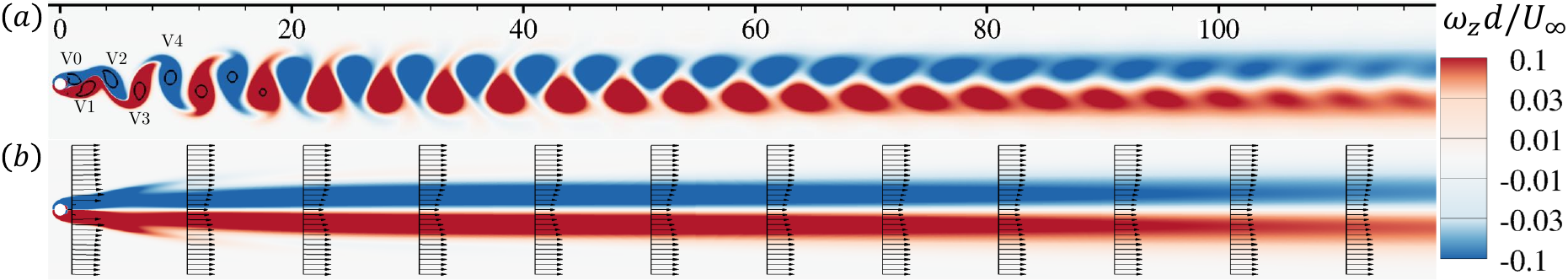}}
	\caption{Vorticity field in the flow past a circular cylinder at $Re=100$. (a) Instantaneous flow at the instant of maximum lift. Black lines are contour lines of $Q/(U_\infty^2d^{-2})=0.1$. V0, V2 and V4 label the most recently formed vortices from the upper surface. V1 and V3 are the most recently formed vortices from the lower surface. (b) Time-averaged flow. Black lines with arrows show the velocity profiles.}
	\label{fig:vorticityfield}
\end{figure}
The numerical simulation is conducted using a high-order spectral/{\it hp} element method, implemented in the open-source code, Nektar++ \citep{cantwell2015nektar++}. The computational domain is $-100d\le x\le 250d$, $-100d\le y\le 100d$, with mesh refinement in the subdomain $-5d\le x \le 160d$, $-5d\le y\le 5d$. The minimum element size on the cylinder surface is $0.01d\times 0.02d$, and the maximum element size in the wake is $0.2d\times0.2d$. The centre of the cylinder is located at $x=0, y=0$. The computational domain contains 57\,009 elements, with a complete polynomial expansion up to the sixth degree applied within each element. The second-order velocity-correction scheme \citep{karniadakis1991high} with a time step of $0.00024 d/U_\infty$ is utilised for the time integration. According to previous studies \citep{gao2019passing,gao2023three}, this spatial and temporal resolution suffices to fully resolve the cylinder wake in the range of $x\le 160d$. The flow is simulated for a total duration of $500 d/U_\infty$ to achieve a periodic flow state. With the flow period denoted by $T$, the dimensionless vortex-shedding frequency is given by $d/(T U_\infty)=0.1636$. Validation of the numerical results is provided in Appendix \ref{app:validationsimu}.

The time-dependent drag $F_x$ and lift $F_y$, non-dimensionalised by $\rho U_\infty^2d$, are presented in figure~\ref{fig:timeforce}. Here, the dimensionless force, rather than the force coefficient, is adopted to facilitate direct comparison with the control-volume integral results. The dimensionless drag fluctuates within a narrow range of $[0.6535, 0.6623]$, with a mean value of 0.6579, whereas the dimensionless lift varies from -0.1587 to 0.1587, with a mean value of 0. The pressure force, indicated by the red solid line, and the friction force, shown by the black dashed line, are also provided in figure~\ref{fig:timeforce}. Whilst the pressure force dominates, the friction force remains non-negligible, accounting for 26\% of the total drag and 12\% of the maximum lift.

The instantaneous vorticity field $\omega_z$, at the moment of maximum lift, is shown in figure~\ref{fig:vorticityfield}(a). At this instant, a vortex, identified by the $Q$-criterion and labelled V0, has just formed from the upper surface of the cylinder. The wake exhibits a typical B\'ernad-von K\'arm\'an vortex street, with concentrated vortices significantly weakened for $x>60d$ under the influence of viscosity. The time-averaged vorticity field is provided in figure~\ref{fig:vorticityfield}(b), alongside mean velocity profiles, which are symmetric about the $x$-axis. A recirculation region extends to $x=1.925d$. The near-wake region ($x<10d$) displays relatively strong nonlinearities, whereas the farther downstream region shows a gradual decay of vorticity. The velocity defect on the centreline persists: at $x=100d$, the centreline velocity is $0.7780 U_\infty$.

Unless otherwise specified, the control volume $V_f$ in the following analyses is taken as an Eulerian domain $-5 d\le x \le x_{\text{wake}}$, $-5d\le y\le 5d$ and $x^2+y^2\ge 0.25d^2$ in the body frame, with $x_{\text{wake}}$ varying from $0.8d$ to $150d$. Because the disturbance velocity and pressure decay rapidly in the upstream direction and in the upward and downward directions, the surface integral over $\Sigma$ can be reduced to a wake-plane integral along the vertical line $-5d \le y\le 5d$ at $x=x_{\text{wake}}$.

\subsection{Total drag from kinetic energy balance}
Analysis of the kinetic energy budget of the disturbance flow provides insight into the power input, energy flux and dissipation in the flow system. By setting the projected weight function $\pW\cdot\pe_x$ as the disturbance velocity $\pu'\coloneq \pu-U_\infty\pe_x$, which becomes zero at infinity and equals $-U_\infty\pe_x$ on the cylinder surface, the kinetic energy balance relation is obtained from equation~\eqref{eq:weightedproject} as 
\begin{equation}\label{eq:kineticEnergy}
\begin{split}
  \underbrace{U_\infty F_x}_{\text{Input power}} &=  \underbrace{\frac{\rd}{\rd t}\int_{V_{f}}{\frac{1}{2}\rho|\pu'|^2 \rd V}}_{\text{Kinetic energy increase}}
  + \underbrace{\int_{\text{wake}}{\frac{1}{2}\rho|\pu'|^2 u_x\rd S}}_{\text{Kinetic energy flux}} 
  + \underbrace{\int_{V_f}{\pT:\bnabla\pu'\rd V}}_{\text{Dissipation rate}} \\
  & +\underbrace{\int_{\text{wake}}{p u'_x \rd S}}_{\text{Pressure power}} 
  +\underbrace{\int_{\text{wake}}{-\ptau\cdot\pu' \rd S}}_{\text{Viscous stress power}} .
\end{split}
\end{equation}
On the wake plane, the volume flux is $\pu\cdot\pe_x=u_x$ and the streamwise disturbance velocity is $\pu'\cdot\pe_x=u'_x$. As the lift force is perpendicular to the cylinder's velocity and thus consumes no power, only the drag force is yielded. Equation~\eqref{eq:kineticEnergy} indicates that the power supplied to the disturbance flow to overcome cylinder drag is partially converted into kinetic energy and dissipated into heat within the control volume $V_f$. The remainder leaves the control volume via the kinetic energy flux through the wake plane, as well as the power exerted by the stress on the wake plane.

\begin{figure}
    \centerline{\includegraphics[width=1\textwidth]{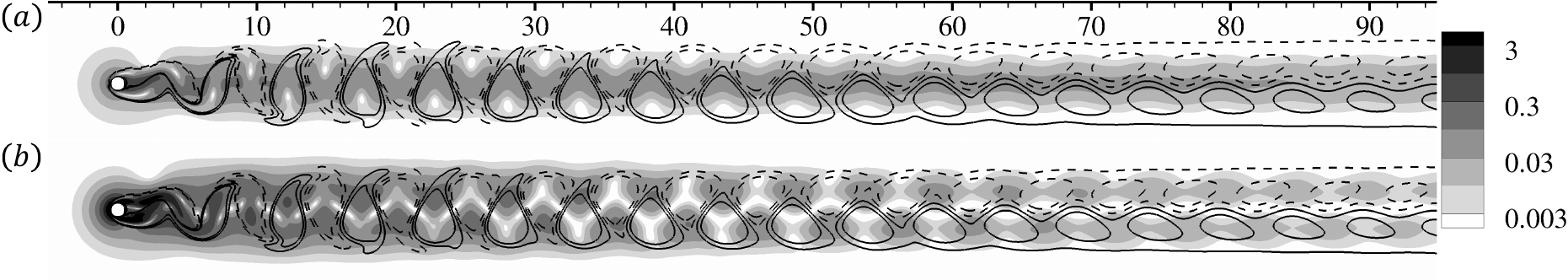}}
	\caption{Instantaneous kinetic energy of the disturbance flow and rescaled dissipation rate at the instant of maximum lift. (a) Kinetic energy, $0.5\rho|\pu'|^2/(\rho U_\infty^2)$. (b) Rescaled dissipation rate, $\pD:\pD/(U_\infty^2d^{-2})$. Black contour lines represent dimensionless vorticity, with negative values (-0.1 and -0.03) as dashed lines and positive values (0.03 and 0.1) as solid lines. }
	\label{fig:kineticdisspation}
\end{figure}

Figure~\ref{fig:kineticdisspation} shows the distribution of the kinetic energy, $0.5\rho|\pu'|^2$, and the dissipation rate, $2\mu\pD:\pD$, of the disturbance flow. The dissipation rate is rescaled to $\pD:\pD=D_{ij}D_{ij}$ due to the small viscosity coefficient $\mu$. Both the kinetic energy and the dissipation rate are mainly distributed in the near wake, and their strengths decrease gradually with the downstream location. The kinetic energy attains its local maximum value around the centre line, i.e. the $x$-axis, whereas the dissipation rate reaches local maximum values in the surroundings of concentrated vortices. In the vortex centre, where the motion of fluid elements is dominated by rigid rotation, the dissipation rate is relatively weak.

\begin{figure}
\centerline{\includegraphics[width=1\textwidth]{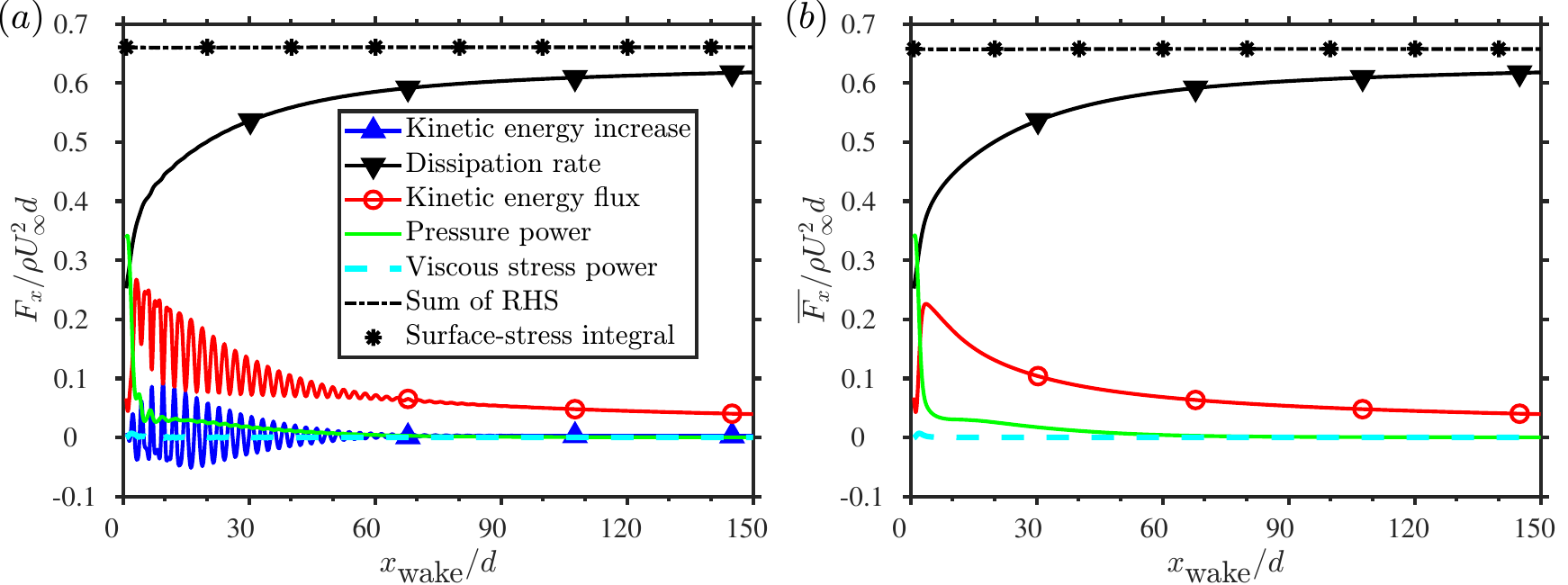}}
	\caption{Drag decomposition based on the kinetic energy balance. (a) Instantaneous drag at the instant of maximum lift. (b) Time-averaged drag.}
	\label{fig:kineticbalance}
\end{figure}

Figure~\ref{fig:kineticbalance}(a) shows the variation of terms in equation~\eqref{eq:kineticEnergy} with respect to the wake-plane location, $x_{\text{wake}}$. The total drag calculated by the control-volume integral \eqref{eq:kineticEnergy}, as indicated by the black dash-dotted line, agrees well with the surface-stress integral shown by the star symbol `*', with a maximum relative difference of $0.2\%$ for $0.8d \le x_{\text{wake}} < 47d$ and $0.03\%$ for $47d\le x_{\text{wake}}\le 150d$. The slightly larger error in the near wake is due to the insufficiency of the wake-plane size. The dissipation-rate term (black line with inverted triangles) dominates the right-hand side (RHS) of equation~\eqref{eq:kineticEnergy} and it grows monotonically with $x_{\text{wake}}$ because its integrand $\pT:\bnabla\pu'=2\mu \pD:\pD$ is always non-negative. The kinetic energy flux is also persistent due to the slow decay of the disturbance velocity in the wake, and it has a value of $0.08F_x U_\infty$ at the downstream location of $x_{\text{wake}}=100d$. Additionally, both the kinetic energy flux (red line with circles) and the kinetic energy increase (blue line with triangles) exhibit significant oscillations when the wake plane cuts vortices in the range of $x_{\text{wake}}<60d$. The pressure power on the wake plane drops quickly and becomes less than $0.05F_x U_\infty$ for $x_{\text{wake}}>7d$. The viscous stress power on the wake plane remains negligible (less than $0.01F_x U_\infty$) throughout the wake range examined.

To eliminate the unsteady effect, the time-averaged form of equation~\eqref{eq:kineticEnergy} is derived as
\begin{equation}\label{eq:meankineticEnergy}
  U_\infty \overline{F_x} =    \int_{\text{wake}}{\frac{1}{2}\rho\overline{|\pu'|^2 u_x}\rd S}  +
  \int_{V_f}{\overline{\pT:\bnabla\pu'}\rd V} + 
  \int_{\text{wake}}{\overline{p u'_x} \rd S} + 
  \int_{\text{wake}}{-\overline{\ptau\cdot\pu'} \rd S}.
\end{equation}
Since the flow is periodic, the time-averaged kinetic-energy-increase term vanishes. As shown in figure~\ref{fig:kineticbalance}(b), terms in equation~\eqref{eq:meankineticEnergy} exhibit similar trends to the instantaneous terms. However, oscillations disappear for the kinetic energy flux, which first grows due to the increase in the advection velocity, $u_x$, and then decreases because of the decay of kinetic energy, $0.5\rho|\pu'|^2$.

In conclusion, the power input that overcomes the drag force of the moving cylinder is mainly dissipated into heat. A small but not negligible portion of power is advected far downstream via the kinetic energy flux. In the near wake of $x_{\text{wake}}<7d$, the power transferred downstream by the pressure stress is also significant. 

\subsection{Total force from the advection-form weighted DMT method}
The advection-form DMT method is characterised by the explicit decomposition of the vortex force. Since the concept of vortex force arises from steady aerodynamics \citep{prandtl1918tragfl} where the velocity employed is the relative velocity between the incoming flow and the body, it is natural to fix the frame of reference on the stationary cylinder, which yields the fluid velocity $\pu$. By setting $W_{ij}=\delta_{ij}$ in equation~\eqref{eq:AdvectionDMT}, the original advection-form DMT method is obtained as
\begin{equation}\label{eq:OriginAdvectionDMT}
\begin{split}
 \pF &=  \underbrace{\int_{V_f}{\rho\pu\times\po \rd V}}_{\text{Vortex force}} +\underbrace{\int_{V_f}{-\rho\px\times\frac{\partial\po}{\partial t} \rd V}}_{\partial_t\po \text{ term}}+\underbrace{\oint_{\partial B}{-\rho\px\times(\pn\times\ppl) \rd S}+ \pF_{\text{DMT},B}}_{\text{Body-surface term}}\\
 &  +\underbrace{\int_{\text{wake}}{-\rho\px\times(\pn\times\ppl) \rd S}}_{\text{Wake-plane integral of $-\rho\px\times(\pn\times\ppl)$}} + \pF_{\text{DMT},\Sigma}.
\end{split}
\end{equation}
The force expression consists of the volume integration of the vortex force, the negative Eulerian time derivative of vorticity momentum, the body-surface term, the wake-plane integral of $-\rho\px\times(\pn\times\ppl)$ and $\pF_{\text{DMT},\Sigma}$. Here, $\pF_{\text{DMT},B}$ and $\pF_{\text{DMT},\Sigma}$ denote $\pF_{w,\text{DMT},B}$ and $\pF_{w,\text{DMT},\Sigma}$ with $W_{ij}=\delta_{ij}$, respectively. The Eulerian time derivative is abbreviated to $\partial_t$. Because $\pu$, as well as $\ppl$ and $\pa$, is zero on the cylinder surface, the body-surface term equals zero. Time averaging can also be conducted on equation~\eqref{eq:OriginAdvectionDMT}, which eliminates the volume integral of $-\rho\px\times \partial\po/\partial t$.

\begin{figure}
\centerline{\includegraphics[width=1\textwidth]{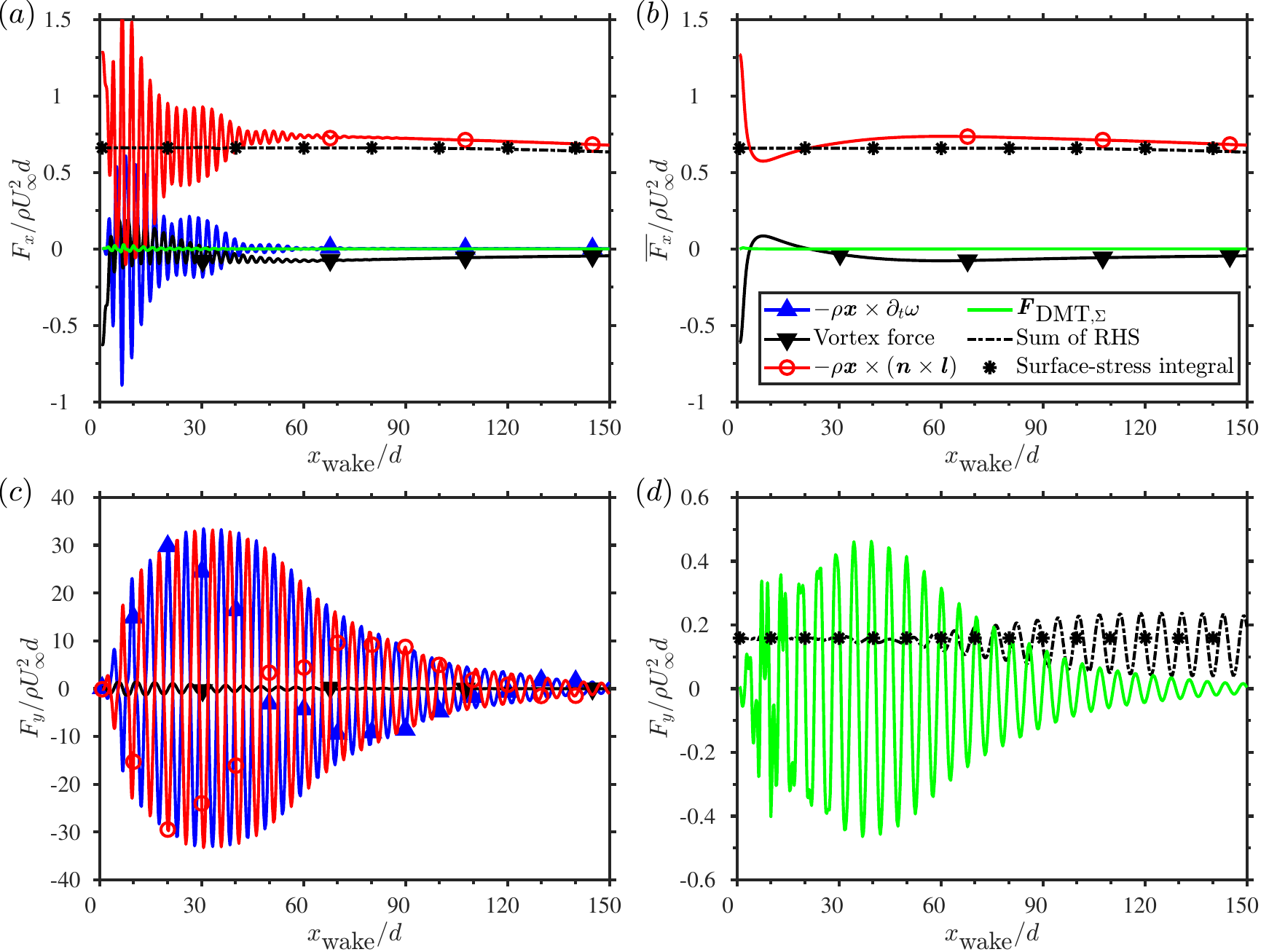}}
	\caption{Force decomposition based on the advection-form DMT method. (a) Instantaneous drag at the instant of maximum lift. (b) Time-averaged drag. (c, d) Instantaneous lift at the instant of maximum lift.}
	\label{fig:DMT}
\end{figure}

Drag decomposition based on equation~\eqref{eq:OriginAdvectionDMT} is shown in figure~\ref{fig:DMT}(a,b). The maximum relative difference between the control-volume integral (black dash-dotted line) and the surface-stress integral (black star symbol) is $0.1\%$ for $x_{\text{wake}}\le 26d$, $1\%$ for $26d<x_{\text{wake}}\le 111d$, and $4\%$ for $111d<x_{\text{wake}}\le 150d$. Interestingly, neither the vortex force nor the volume integral of $-\rho\px\times\partial\po/\partial t$ dominates the drag. The vortex force even contributes a small negative drag after experiencing dramatic oscillations in the range of $x_{\text{wake}}<60d$. Instead, the wake-plane integral of $-\rho\px\times(\pn\times\ppl)$ is much closer to the total drag, although it also exhibits oscillations for $x_{\text{wake}}<60d$. Throughout the $x_{\text{wake}}$ range examined, $\pF_{\text{DMT},\Sigma}$ is negligibly small. The time-averaged curves (figure~\ref{fig:DMT}b) show that the vortex force first has a large negative value and increases rapidly to a slightly positive value, before slowly decreasing to a small negative plateau value. The time-averaged wake-plane integral of $-\rho\px\times(\pn\times\ppl)$ exhibits a contrary trend to the time-averaged vortex force. 

Lift decomposition based on equation~\eqref{eq:OriginAdvectionDMT} is shown in figure~\ref{fig:DMT}(c,d). Due to the large value of $x$, the volume integral of $-\rho(\px\times\partial\po/\partial t)\cdot\pe_y=\rho x\partial\omega_z/\partial t$ and wake-plane integral of $-\rho\left[\px\times(\pn\times\ppl)\right]\cdot\pe_y=\rho x l_y$ exhibit violent oscillations, with an increasing amplitude for $x_{\text{wake}}<30d$ and a decreasing amplitude for $x_{\text{wake}}>30d$. Significant oscillations also exist in the vortex-force term and the $\pF_{\text{DMT},\Sigma}$ term. The relative error of the control-volume integral~\eqref{eq:OriginAdvectionDMT}, compared with the surface-stress integral, is large at far downstream. In the range of $x_{\text{wake}}\le 3.8d$, this relative error is less than $1\%$; it quickly grows to $5.3\%$ at $x_{\text{wake}}=9.5d$ and to $10\%$ at $x_{\text{wake}}= 35d$. At $x_{\text{wake}}=98d$, this relative error can be as high as $60\%$, indicating deterioration in accuracy due to the large control volume.

To overcome the issue caused by the large $x$ value in equation~\eqref{eq:OriginAdvectionDMT}, the divergence- and curl-free auxiliary function introduced in the work of \citet{quartapelle1983force} is adopted but with a negative sign. Here, we set $\pW=\bnabla\pphi = \bnabla\phi_x\pe_x+\bnabla\phi_y\pe_y$. In the fluid domain, $\pphi$ is governed by the Laplacian equation,
\begin{equation}\label{eq:philaplace}
    \nabla^2\pphi = \pzero.
\end{equation}
On the cylinder surface, $\pphi$ satisfies the following Neumann boundary condition
\begin{equation}
\label{eq:phibc123}    \hat{\pn}\cdot\bnabla\pphi =\hat{\pn},
\end{equation}
and $\pphi$ goes to zero at the infinitely far field. $\bnabla\phi_x$ or $\bnabla\phi_y$ can be seen as the potential flow generated by the motion of the body in the $x$ or $y$ direction with a unit speed. It is noted that $\bnabla\pphi$ is dimensionless and that $\pphi$ has a dimension of length.
At the far field, $\bnabla\pphi$ decreases to zero at a rate of $|\px|^{-2}$, which effectively reduces the magnitudes of integrands. The advection-form weighted DMT method is written as
\begin{equation}\label{eq:GradPhiAdvectionDMT}
\begin{split}
 \pF &=   \underbrace{\int_{V_f}{\rho(\pu\times\po)\cdot\bnabla\pphi \rd V}}_{\text{Weighted vortex force}} +\underbrace{\int_{V_f}{-\rho\px\times(\bnabla\times\pb_w) \rd V}}_{\pb_w\text{ term}} \\
 &  +\underbrace{\oint_{\partial B}{-\rho\px\times(\pn\times\ppl_w) \rd S} + \pF_{w,\text{DMT},B} + \left(\pF_f - \pF_{w,f}\right)}_{\text{Body-surface term}} \\
 & + \underbrace{\int_{\text{wake}}{-\rho\px\times(\pn\times\ppl_w) \rd S}}_{\text{Wake-plane integral of $-\rho\px\times(\pn\times\ppl_w)$}} + \pF_{w,\text{DMT},\Sigma}.
\end{split}
\end{equation}
The force expression consists of the volume integral of the weighted vortex force, the volume integral of $-\rho\px\times(\bnabla\times\pb_w)$, the body-surface term, the wake-plane integral of $-\px\times(\pn\times\ppl_w)$ and $\pF_{w,\text{DMT},\Sigma}$. It is noted that the weighted vortex force is exactly the volume force element in the work of \citet{chang1992potential} and \citet{chiu2023vorticity}. The $\pb_w$ term contains not only the unsteadiness effect but also spatial derivatives of the weight function; as a result, the time averaging of $-\rho\px\times(\bnabla\times\pb_w)$ is not zero. $\pf_w$ is not zero on the cylinder surface, either, and hence $\pF_{w,\text{DMT},B}$ also contributes to the force. Additionally, although $\pF_{w,p}$ equals $\pF_p$, $\pF_{w,f}$ does not equal $\pF_f$. Therefore, a term $\pF_f-\pF_{w,f}$ is introduced in the body-surface term.

\begin{figure}
\centerline{\includegraphics[width=1\textwidth]{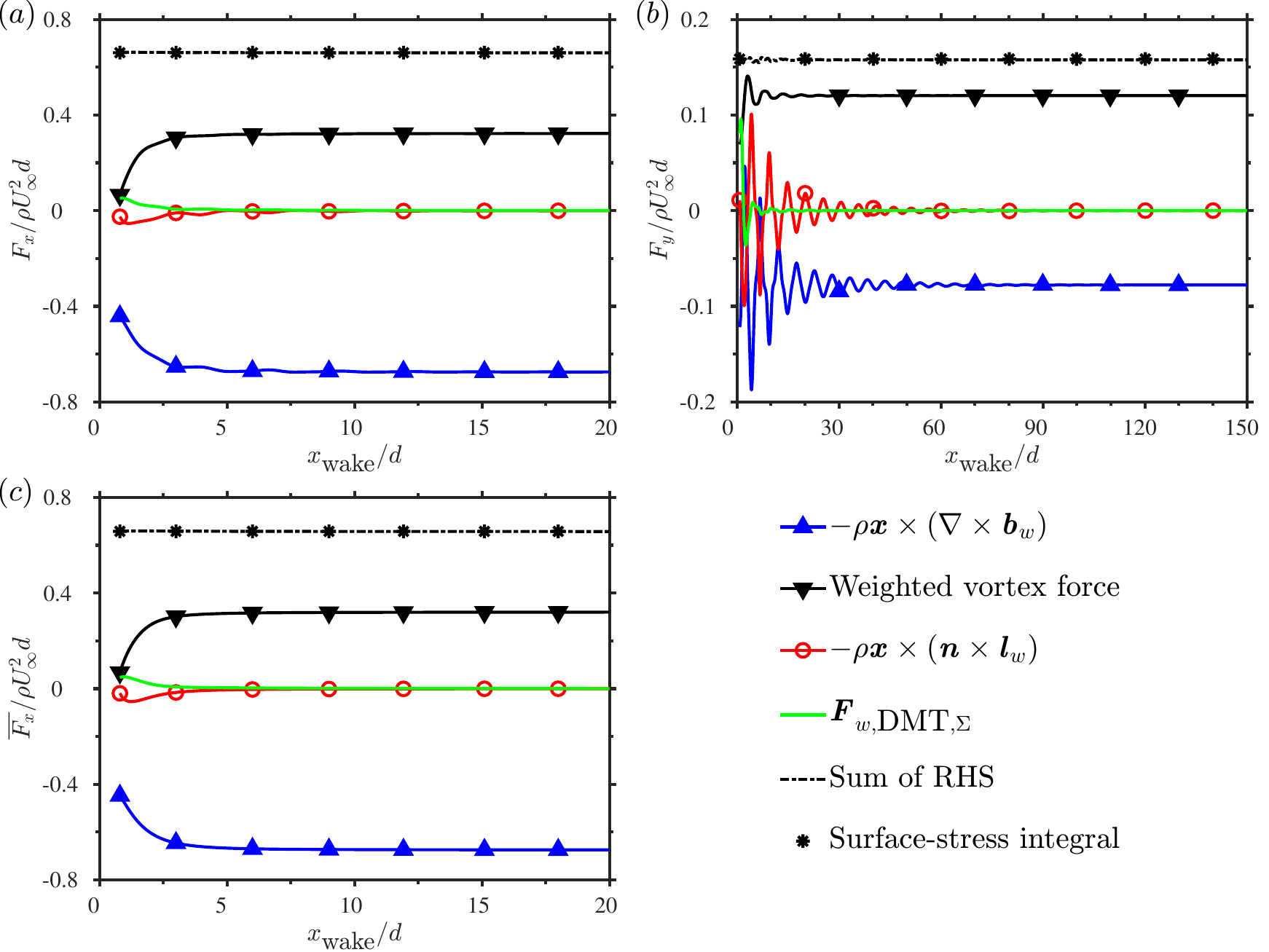}}
	\caption{Force decomposition based on the advection-form weighted DMT method. (a) Instantaneous drag and (b) instantaneous lift at the instant of maximum lift. (c) Time-averaged drag. }
	\label{fig:DMTweight}
\end{figure}

Figure~\ref{fig:DMTweight}(a,b) shows the force decomposition at the maximum-lift instant based on the advection-form weighted DMT method~\eqref{eq:GradPhiAdvectionDMT}. The curves converge to constant values quickly with the increasing $x_{\text{wake}}$. Specifically, the drag components (figure~\ref{fig:DMTweight}a) remain almost invariant for $x_{\text{wake}}>10d$. The lift components (figure~\ref{fig:DMTweight}b), which contain $x$, show damped oscillations and reach almost constant values for $x_{\text{wake}}>70d$. The wake-plane integrals (red line with circle and green line) both decay to zero. The numerical accuracy of the control-volume integral (black dash-dotted line) is also significantly improved through the introduction of the weight function. The maximum relative error in the instantaneous drag is $0.28\%$ for $x_{\text{wake}}< 5d$ and drops below $0.07\%$ for $5d\le x_{\text{wake}}\le 150d$. The maximum relative error in the lift is slightly larger, which is $2.6\%$ for $x_{\text{wake}}\le 28d$ and $0.7\%$ for $28d<x_{\text{wake}}\le 150d$.

In the weighted DMT method~\eqref{eq:GradPhiAdvectionDMT}, the volume integral of the weighted vortex force and the body-surface term become the main contributions to both the drag and the lift. The volume integral of $-\rho\px\times(\bnabla\times\pb_w)$, on the other hand, has a strong negative force contribution. When the final constant values are reached, the body surface term equals $1.53F_x$ and $0.73F_y$, the volume integral of the weighted vortex force equals $0.49F_x$ and $0.76F_y$, and the volume integral of $-\rho\px\times(\bnabla\times\pb_w)$ equals $-1.02F_x$ and $-0.49F_y$, at the maximum-lift instant. It is also noted that the instantaneous drag components are almost the same as the time-averaged drag components (figure~\ref{fig:DMTweight}c), which is consistent with the small oscillation amplitude of the time-dependent drag.
\begin{figure}
\centerline{\includegraphics[width=1\textwidth]{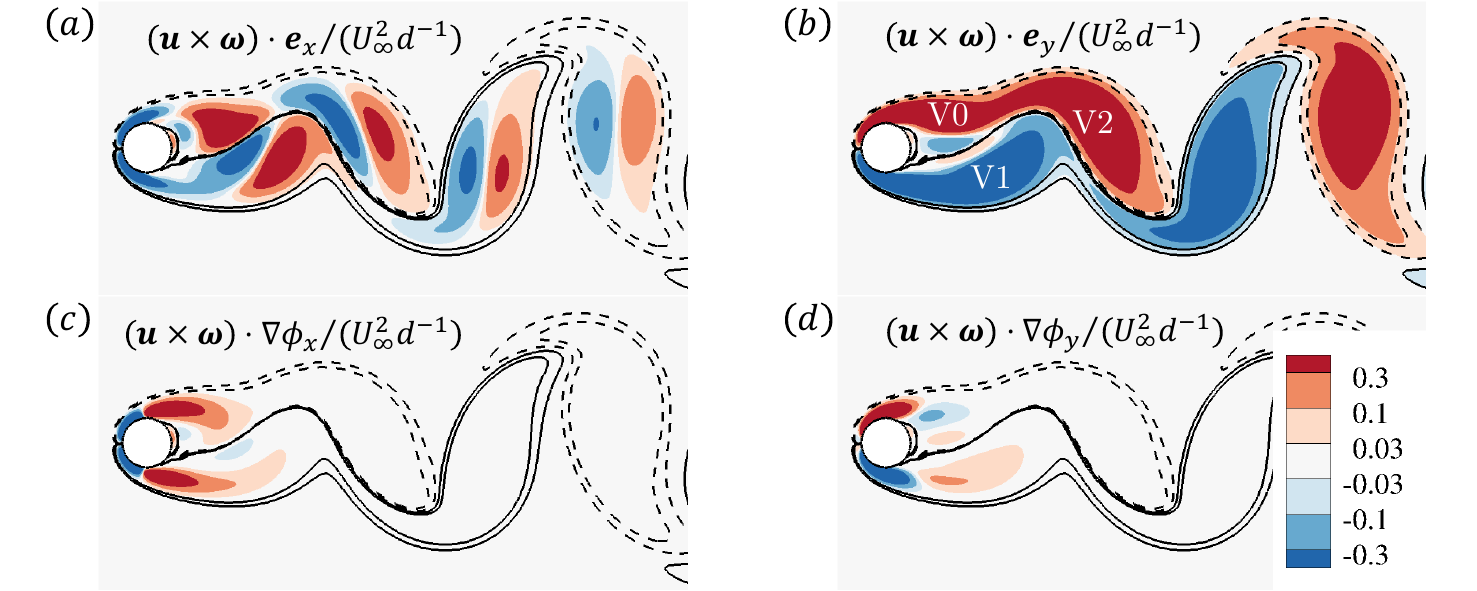}}
	\caption{Colour shows the distribution of (a,b) vortex force and (c,d) weighted vortex force at the instant of maximum lift. Black contour lines represent dimensionless vorticity, with negative values (-0.1 and -0.03) as dashed lines and positive values (0.03 and 0.1) as solid lines.}
	\label{fig:vortexforce}
\end{figure}

The spatial distribution of the vortex force is shown in figure~\ref{fig:vortexforce}(a,b) to gain insights into its integrated behaviour presented in figure~\ref{fig:DMT}. The vorticity field is also shown by black contour lines with positive values indicated by solid lines and negative values in dashed lines. The three most recently formed vortices are labelled as V0, V1 and V2 in figure~\ref{fig:vortexforce}(b). By definition, the vortex force is distributed in high-vorticity regions, such as the boundary layers, free shear layers and wake vortices. Within each concentrated vortex except for V0, the $x$-component of the vortex force $u_y\omega_z$ is negative on the left side and positive on the right side, owing to the opposite signs of the vertical velocity across the vortex centre. In contrast, the $y$-component of the vortex force $-u_x\omega_z$ exhibits a distribution similar to that of the negative vorticity field, because $u_x$ is generally positive outside the recirculation region. Due to the slow decay of concentrated vortices in the wake, the volume integral of vortex force inevitably oscillates when the wake plane passes through vortices, and the oscillation frequency of the $x$-component is twice that of the $y$-component, as indicated by the black lines with inverted triangles in figure~\ref{fig:DMT}(a,c).

The spatial distribution of the weighted vortex force is shown in figure~\ref{fig:vortexforce}(c,d).
In contrast to the original vortex force, the weighted vortex force is mainly distributed in the boundary layers and the free shear layers within the range $x<3d$. Its magnitude is small even in the concentrated vortex V2. Because $\phi_x=-0.25 d^2 x/(x^2+y^2)$ changes sign across the $y$ axis, the $x$-component of the weighted vortex force is negative in a small area in the boundary layer of the left-half-cylinder, whereas it becomes positive in larger areas in the shear layers and vortices with $x>0$; as a result, its volume integral increases monotonically to a positive value of $0.32\rho U_\infty^2 d$ in the range $0.8d\le x_{\text{wake}}<5d$ (see the black line with inverted triangles in figure~\ref{fig:DMTweight}a). Similarly, $\phi_y=-0.25 d^2 y/(x^2+y^2)$ has different signs across the $x$ axis; therefore, the $y$-component of the weighted vortex force is positive in the boundary layer of the upper surface, while it becomes negative in the boundary layer of the lower surface. Additionally, the $y$-component of the weighted vortex force is negative in vortex V0, whereas it has a relatively strong positive value in vortex V1; as a result, its volume integral grows to a positive value in the range $0.8d<x_{\text{wake}}<2.9d$ before exhibiting small-amplitude and damped oscillations for $x_{\text{wake}}>2.9d$, see the black line with inverted triangles in figure~\ref{fig:DMTweight}(b).

In conclusion, the advection-form weighted DMT method~\eqref{eq:GradPhiAdvectionDMT} recovers the volume force element proposed by \citet{chang1992potential} when the negative gradient of the auxiliary potential introduced by \citet{quartapelle1983force} is employed as the weight function. The introducing of a rapidly decaying weight function offers advantages in improving the numerical accuracy and the control-volume convergence of the DMT method.
\subsection{Pressure distribution over the cylinder surface}
\label{subsec:pdistribution}
A quantitative relationship between the pressure distribution over the cylinder surface and flow structures is established and evaluated in this section. A series of harmonic functions that satisfy boundary conditions
\begin{equation}
    \hat{\pn}\cdot\bnabla\varphi_{cm} = \cos(m\theta) ,\quad
    \hat{\pn}\cdot\bnabla\varphi_{sm} = \sin(m\theta), \quad m=0,1,2,...
\end{equation}
on the cylinder surface, with $\theta$ being the polar angle, is considered. The corresponding weighted pressure forces are
\begin{subequations}\label{eq:wallFcsmp}
\begin{align}
 \label{eq:wallFcmp}   F_{cm,p}(t)  &\coloneq \oint_{\partial B}{ -p\hat{\pn}\cdot\bnabla\varphi_{cm}\rd S} =  -0.5 d \int_0^{2\pi}{p\cos(m\theta)\rd \theta},\\
 \label{eq:wallFsmp}   F_{sm,p}(t)  &\coloneq \oint_{\partial B}{ -p\hat{\pn}\cdot\bnabla\varphi_{sm}\rd S} = -0.5 d \int_0^{2\pi}{p\sin(m\theta)\rd \theta}.
\end{align}
\end{subequations}
The $F_{cm,p}$ and $F_{sm,p}$, which are termed pressure distribution coefficients, are proportional to, but with opposite signs to, the Fourier expansion coefficients of the surface pressure. By definition, $F_{s0,p}=0$. It is also noted that the total pressure force is $\pF_p=F_{c1,p}\pe_x+ F_{s1,p}\pe_y$. The Fourier expansion of the surface pressure is given by
\begin{equation}
  \label{eq:pFourierexp}  p_N(t, \theta) = -\frac{1}{\pi d}F_{c0,p}(t)-\frac{2}{\pi d}\sum_{m=1}^{N}\left[F_{cm,p}(t)\cos(m\theta)+F_{sm,p}\sin(m\theta)\right],
\end{equation}
which recovers the surface pressure distribution when the expansion order $N$ goes to infinity,
\begin{equation}
    p(t, \theta) = \lim_{N\to+\infty}{p_N(t, \theta)}.
\end{equation}

Since $\nabla^2\varphi_{cm}=0$ and $\nabla^2\varphi_{sm}=0$, $F_{cm,p}$ and $F_{sm,p}$ can be expressed as control-volume integrals through the weighted projection method with curl-free weight functions~\eqref{eq:potentialprojectvector}. These are given by
\begin{subequations}\label{eq:volFcsmp}
\begin{align}
\label{eq:volFcmp} F_{cm,p} & =  \underbrace{\int_{V_{f}}{-2\rho Q \varphi_{cm}  \rd V}}_{\text{Weighted integral of $Q$}} + \underbrace{\oint_{\partial B}{\pn\cdot(-\rho\pa+\mu\nabla^2\pu)\varphi_{cm} \rd S}}_{\text{Body surface term}} \\
\nonumber & + \underbrace{\int_{\text{wake}}{\pn\cdot(-\rho\pa+\mu\nabla^2\pu)\varphi_{cm} \rd S} + \int_{\text{wake}}{-p\pn\cdot\bnabla\varphi_{cm} \rd S}}_{\text{Wake plane integral}},\\
\label{eq:volFsmp} F_{sm,p} & =  \int_{V_{f}}{-2\rho Q \varphi_{sm}  \rd V} + \oint_{\partial B}{\pn\cdot(-\rho\pa+\mu\nabla^2\pu)\varphi_{sm} \rd S} \\
\nonumber & + \int_{\text{wake}}{\pn\cdot(-\rho\pa+\mu\nabla^2\pu)\varphi_{sm} \rd S} + \int_{\text{wake}}{-p\pn\cdot\bnabla\varphi_{sm} \rd S}.
\end{align}
\end{subequations}
Therefore, equations~\eqref{eq:pFourierexp} and \eqref{eq:volFcsmp} establish a quantitative relationship between the $Q$ field and the surface pressure distribution.

An alternative interpretation of equation~\eqref{eq:volFcmp}, as well as \eqref{eq:volFsmp}, is given in Appendix~\ref{app:pressurepartition}. The pressure field can be decomposed into four components, $p=p_Q+p_a+p_{\text{vis}}+p_\infty$. Here, $p_Q$ represents the partial pressure field generated by the volume pressure source $2\rho Q$, $p_a$ is generated by the body-surface normal acceleration, $p_{\text{vis}}$ is generated by the viscosity effect on $\partial B$ and $p_\infty$ is the pressure at infinity. Correspondingly, $F_{cm,p}=F_{cm,p_Q}+F_{cm,p_a}+F_{cm,p_{\text{vis}}}+F_{cm,p_\infty}$. Auxiliary functions $\varphi_{cm}$ and $\varphi_{sm}$ are the solutions to the adjoint equation of the pressure Poisson equation. Therefore, $F_{cm,p_Q}$, the body-surface integral of $-p_Q \cos(m\theta)$, can be calculated by the volume integration of $-2\rho Q\varphi_{cm}$ over the whole fluid domain. Here, the function $\varphi_{cm}$ represents the sensitivity of $F_{cm,p_Q}$ to the volume pressure source. A vortical structure with $Q>0$ produces a low-pressure region through ``vortex suction" on the nearby body surface, whereas the strain rate region with $Q<0$ generates a repulsive force and a high-pressure area on the neighbouring solid surface. Similarly, $F_{cm,p_a}$ and $F_{cm, p_{\text{vis}}}$ can be calculated by the surface integration of $-\rho\pn\cdot\pa\varphi_{cm}$ and $\mu\pn\cdot\nabla^2\pu\varphi_{cm}$ on $\partial B$, respectively. The outer surface integral of $-p\pn\cdot\bnabla\varphi_{c0}$ converges to $\pi dp_\infty$, when $\Sigma$ approaches infinity.

For the circular cylinder, the auxiliary functions $\varphi_{cm}$ and $\varphi_{sm}$ can be solved analytically as
\begin{subequations}
\begin{align}
\varphi_{c0} &= 0.5 d\ln (2r/d),\quad \varphi_{s0} = 0,\\
\varphi_{cm} &= -m^{-1} (d/2)^{m+1} r^{-m}\cos(m\theta),\quad m=1,2,3,... \\
\varphi_{sm}& = -m^{-1} (d/2)^{m+1} r^{-m} \sin(m\theta),\quad m=1,2,3,...
\end{align}
\end{subequations}
Here, $r$ and $\theta$ are the polar coordinates, with $x=r\cos\theta$ and $y=r\sin\theta$. For $m=0$, the net flux of $\bnabla\varphi_{c0}$ from the cylinder surface is non-zero, and therefore $\varphi_{c0}$, which can be seen as the velocity field induced by a point source of strength $\pi d$ in the 2D potential flow, grows as $O(\ln r)$ at the far field. Here, $\varphi_{c0}$ is set to zero on the cylinder surface. For $m\ge1$, $\varphi_{cm}$ and $\varphi_{sm}$ decay as $O(r^{-m})$ when $r$ goes to infinity.

\begin{figure}
\centerline{\includegraphics[width=1\textwidth]{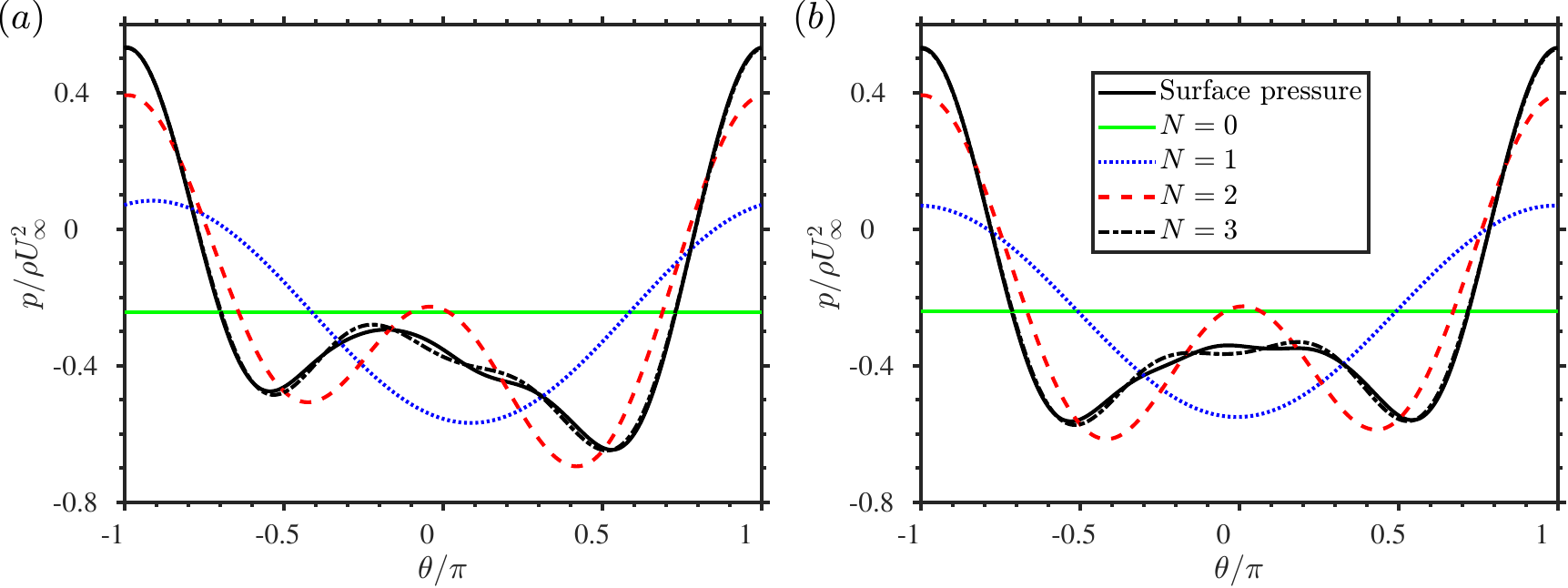}}
	\caption{Pressure distribution on the cylinder surface and its Fourier expansion $p_N$ with orders $N=0,1,2,3$. (a) At the instant of maximum lift. (b) At the instant of zero lift.}
	\label{fig:Fouriersurfp}
\end{figure}

Figure~\ref{fig:Fouriersurfp} shows the pressure distribution over the cylinder surface at two instants. Here, $\theta=\pm\pi$ is the leading edge and $\theta=0$ is the trailing edge of the cylinder. Meanwhile, $\theta=\pi/2$ is the highest point and $\theta=-\pi/2$ is the lowest point on the cylinder surface. The leading-edge stagnation point has a higher pressure than the leeward side of the cylinder, while two local pressure minima occur slightly upstream of the highest and the lowest points. At the maximum-lift instant, the upper-right quarter ($0<\theta<0.5\pi$) has a lower pressure than the lower-right quarter ($-0.5\pi<\theta<0$); whereas at the zero-lift instant, the pressure distribution is more symmetric between the upper and the lower surfaces. 

The truncated Fourier expansion~\eqref{eq:pFourierexp} is also shown in figure~\ref{fig:Fouriersurfp}, with the dimensionless $F_{cm,p}$ and $F_{sm,p}$ at the maximum-lift instant presented in Table~\ref{tab:Fscm}. With the increasing expansion order $N$, the Fourier expansion $p_N$ gradually converges to the surface pressure $p$. The zeroth-order Fourier expansion gives the surface-averaged pressure, which is a constant lower than $p_\infty = 0$ as indicated by the green solid line. However, due to the minus sign in equation~\eqref{eq:wallFcmp}, $F_{c0,p}$ has a positive value of $0.76\rho U_\infty^2 d$. The first-order expansion, as indicated by the blue-dotted line, captures the higher pressure near the leading edge and the lower pressure on the leeward side because of the contribution from $F_{c1,p}$, i.e. the pressure drag. At the maximum-lift instant (figure~\ref{fig:Fouriersurfp}a), the line with $N=1$ exhibits asymmetry about $\theta=0$ owing to the significance of the antisymmetric term $F_{s1,p}=0.1395\rho U_\infty^2 d$, which equals the pressure lift. Since $F_{s2,p}$ and $F_{s3,p}$ are negligibly small, the second-order coefficient $F_{c2,p}$ with a negative value of $-0.5035\rho U_\infty^2 d$ captures the even higher pressure at the leading edge and the local pressure minima near the highest and the lowest points, see the red-dashed line. With $N=3$ as indicated by the black dash-dotted line, the third-order coefficient $F_{c3,p}$ further raises the pressure at the leading edge and reduces $p$ near the trailing edge, resulting in a satisfactory approximation to the surface pressure.

\begin{table}
  \begin{center}
  \begin{tabular}{lccccccccc}
  &  $F_{c0,p}$ & $F_{c1,p}$ & $F_{s1,p}$ & $F_{c2,p}$ & $F_{s2,p}$ & $F_{c3,p}$ & $F_{s3,p}$ & $F_{c4,p}$ & $F_{s4,p}$  \\
 \hline
  Weighted integral of  $Q$ &  0.7609 & 0.3231 & 0.1203 &-0.3335 & 0.0125 & 0.1785 & 0.0026 & -0.0264 & 0.0077 \\
  Body surface term & 0.0000 & 0.1688 & 0.0192 & -0.1700 & 0.0134 & 0.0389 & 0.0049 & 0.0092 & 0.0034\\
  Sum of RHS terms & 0.7611 & 0.4919 & 0.1395 & -0.5035 & 0.0259 & 0.2175 & 0.0075 & -0.0173 & 0.0111 \\
   Surface integral results & 0.7617 & 0.4919 & 0.1395 & -0.5035 & 0.0259 & 0.2175 & 0.0075 & -0.0173 & 0.0111
 \\
  \end{tabular}
  \caption{The decomposition of $F_{cm,p}/(\rho U_\infty^2 d)$ and $F_{sm,p}/(\rho U_\infty^2 d)$ when the integrals are independent of the control volume size. $x_{\text{wake}}=150d$ for $m=0$; $x_{\text{wake}}=36 d$ for $m=1,2,3$, 4.} 
  \label{tab:Fscm}
  \end{center}
\end{table}

The decomposition of dimensionless $F_{cm,p}$ and $F_{sm,p}$ at the maximum-lift instant, based on equation~\eqref{eq:volFcsmp}, is provided in Table~\ref{tab:Fscm}. A large enough control volume, which is $x_{\text{wake}}=150d$ for $m=0$ and $x_{\text{wake}}=36d$ for $m\ge1$, is adopted to eliminate the $x_{\text{wake}}$ dependence. The results calculated by the control-volume integral~\eqref{eq:volFcsmp} and the cylinder surface integral~\eqref{eq:wallFcsmp} are almost identical. The weighted integral of $Q$ over the control volume is the major contribution to the pressure distribution coefficients. Although the cylinder is stationary and thus $\pa=\pzero$, the body surface term with $m\ge1$ is not zero due to the viscosity effect in the Neumann pressure boundary condition on the cylinder surface, $\hat{\pn}\cdot\bnabla p = -\rho\hat{\pn}\cdot\pa+\mu\nabla^2\pu\cdot\hat{\pn}$. For prominent coefficients $F_{c1,p}$ and $F_{c2,p}$, the contribution from the body surface term is about half that of the weighted integral of $Q$, representing a non-negligible viscosity effect on the surface pressure distribution.


\begin{figure}
\includegraphics[height=0.35\textwidth]{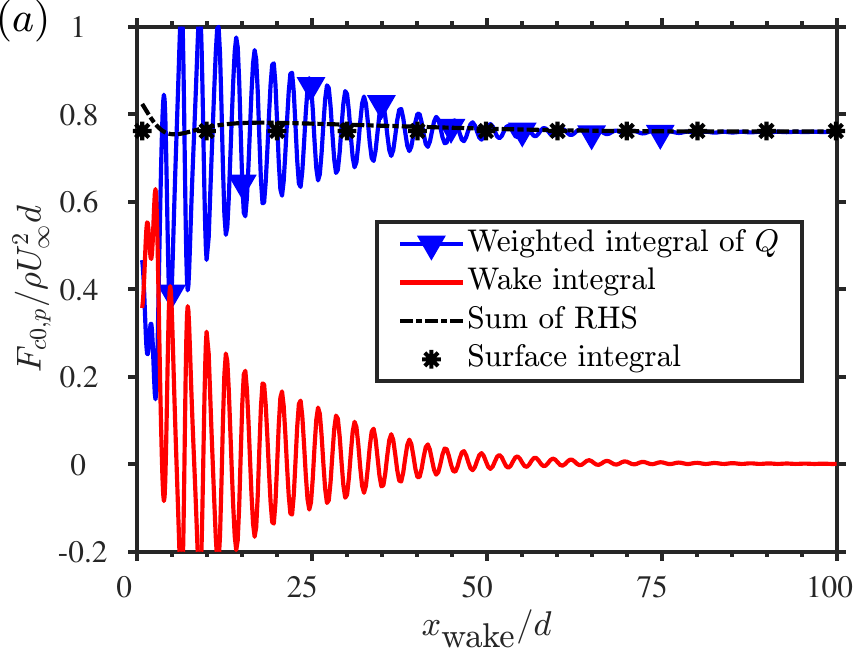}
\raisebox{0.1\textwidth}{
\includegraphics[height=0.25\textwidth]{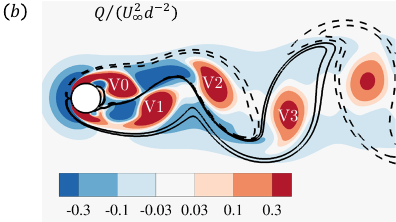}}
\caption{(a) Decomposition of the zeroth-order pressure distribution coefficient, $F_{c0,p}$.  (b) Spatial distribution of $Q$ at the instant of maximum lift. Black contour lines represent dimensionless vorticity, with negative values (-0.1 and -0.03) as dashed lines and positive values (0.03 and 0.1) as solid lines.}
	\label{fig:projWQ0}
\end{figure}

The dependence of components in $F_{c0,p}$ on the wake-plane location $x_{\text{wake}}$ is shown in figure~\ref{fig:projWQ0}(a). The body surface term is zero since $\varphi_{c0}$ is set to zero on the cylinder surface. Because the total flux of $\bnabla\varphi_{c0}$ through $\Sigma$ remains a constant value of $\pi d$, the outer surface integral of $-p\pn\cdot\bnabla\varphi_{c0}$ converges to $-\pi d p_\infty$, with $p_\infty$ set to zero, when $\Sigma$ goes to infinity. 
Since $\varphi_{c0}$ grows with $r$, the decay of $-2\rho Q\varphi_{c0}$ at the far field solely depends on the decrease of $Q$. As a result, the volume integral of $-2\rho Q\varphi_{c0}$ (blue line with inverted triangles) converges slowly. For $0.8d\le x_{\text{wake}}<10d$, whilst experiencing strong oscillations, this volume integral exhibits an increasing trend. Further downstream, it exhibits damped oscillations before reaching a constant value of $0.76\rho U_\infty^2 d$ at $x_{\text{wake}}=75d$.

The spatial distribution of $Q$ is shown in figure~\ref{fig:projWQ0}(b) to gain further insights. Due to the singularity in 2D space, a constant in $\varphi_{c0}$ is undetermined. Although $\varphi_{c0}$ has been artificially set to zero on $\partial B$, the volume integrand $-2\rho Q\varphi_{c0}$ cannot objectively reflect the contribution from local fluid domains. When the total volume integration of $2\rho Q$ is zero, which is satisfied in the present flow, Appendix \ref{app:pressurepartition} provides another explanation for the cause of the average surface pressure: if the region with $Q>0$ is distributed closer to the solid body than the region with $Q<0$, the average surface pressure will become lower than $p_\infty$. Since $Q$ measures the relative strength between rotation and deformation of fluid elements, $Q$ is positive at the centre of concentrated vortices, such as the most recently formed vortices V0, V1 and V2, and in the rolling-up shear layers; whereas it is negative between two successive vortices and in front of the leading edge. Clearly, the rolling-up shear layers near the cylinder and the vortex centre of V0 show strong positive values of $Q$, and they are closest to the cylinder. The strain rate region below V0 is weaker than the vortex centre of V0 in magnitude; the strain rate region between V0 and V2 is also weaker than V1. Therefore, the stronger shear layers and vortices together lead to the lower average pressure on the cylinder surface.

\begin{figure}
\centerline{\includegraphics[width=1\textwidth]{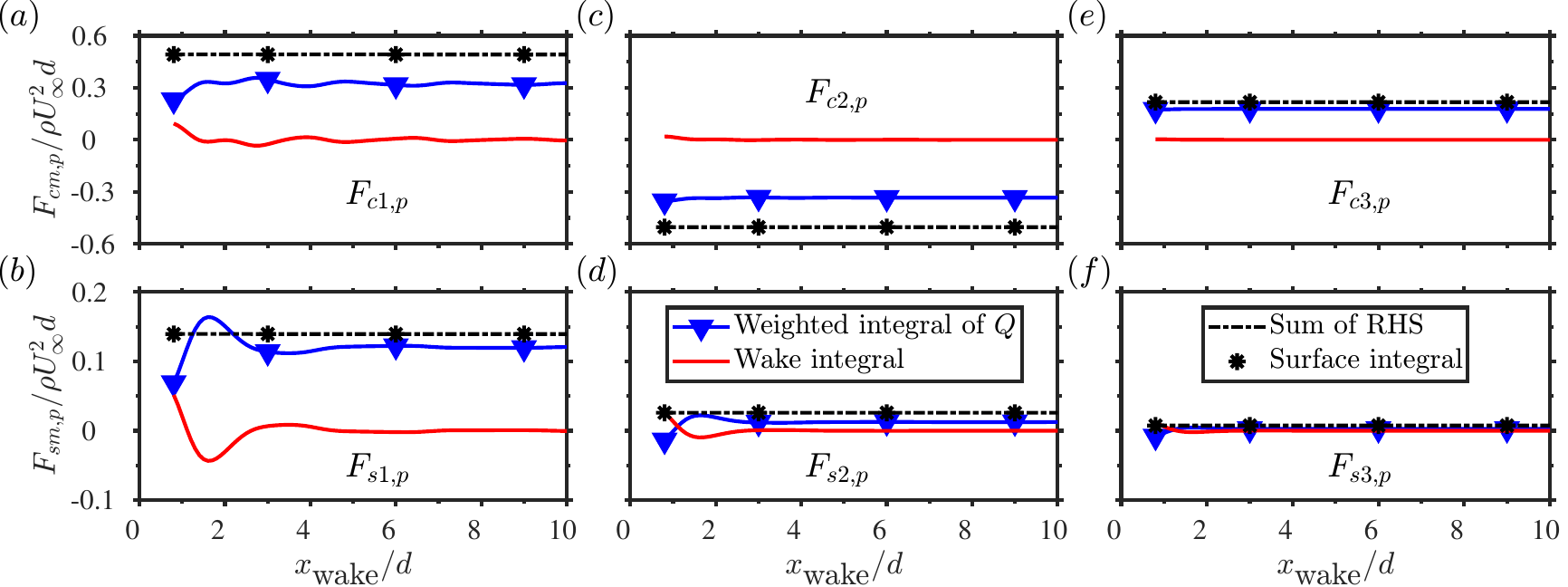}}
	\caption{Decomposition of the pressure distribution coefficients, $F_{cm,p}$ and $F_{sm,p}$ with $m=1,2,3$, based on the weighted projection method with curl-free weight function. Panels (a,c,e) are $F_{cm,p}$. Panels (b,d,f) are $F_{sm,p}$.}
	\label{fig:projFm}
\end{figure}

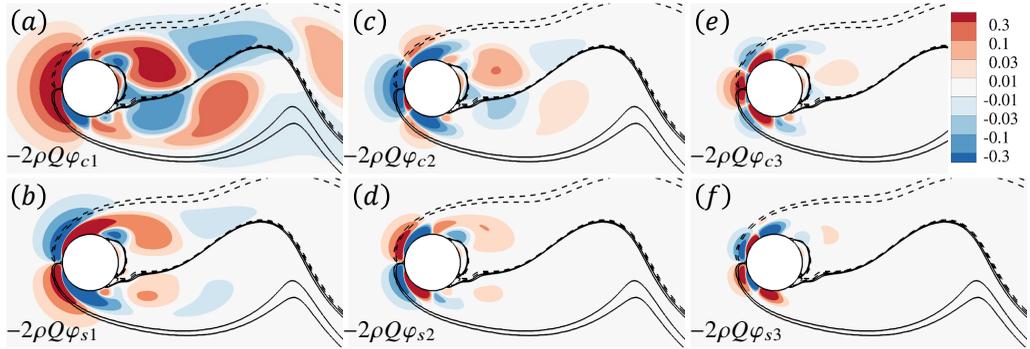
\begin{figure}
\centering
\begin{overpic}[width=1\linewidth]{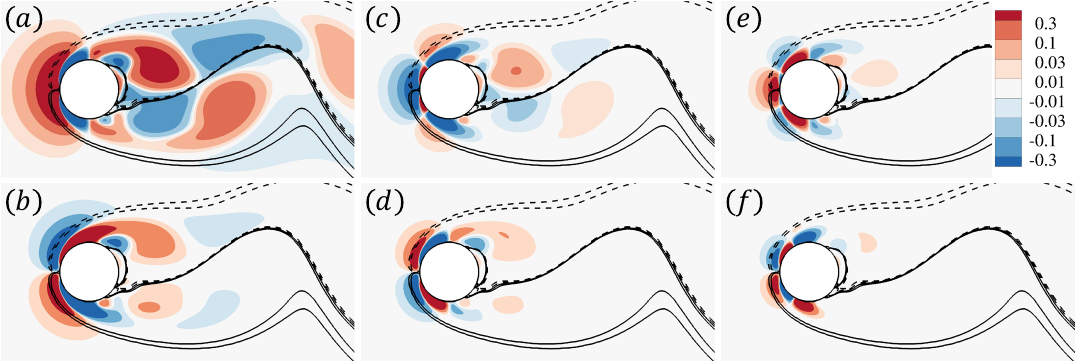}
    \put(0,18){\small $-2\rho Q\varphi_{c1}$}
    \put(33,18){\small $-2\rho Q\varphi_{c2}$}
    \put(67,18){\small $-2\rho Q\varphi_{c3}$}
    \put(0,1){\small $-2\rho Q\varphi_{s1}$}
    \put(33,1){\small $-2\rho Q\varphi_{s2}$}
    \put(67,1){\small $-2\rho Q\varphi_{s3}$}
\end{overpic}
\caption{The spatial distribution of (a, c, e) $-2\rho Q\varphi_{cm}/(\rho U_\infty^2 d^{-1})$ and (b, d, f) $-2\rho Q\varphi_{sm}/(\rho U_\infty^2 d^{-1})$ at the instant of maximum lift. (a, b) $m=1$. (c, d) $m=2$. (e, f) $m=3$. Black contour lines represent dimensionless vorticity, with values (-0.1 and -0.03) as dashed lines and positive values (0.03 and 0.1) as solid lines.}
	\label{fig:projQm}
\end{figure}

The dependence of $F_{cm,p}$ and $F_{sm,p}$ with $m=1,2,3$ on the $x_{\text{wake}}$ is shown in figure~\ref{fig:projFm}. 
For $m\ge 1$, the total fluxes of $\bnabla\varphi_{cm}$ and $\bnabla\varphi_{sm}$ over $\Sigma$ are both zero; therefore, the wake-plane integral (red line) decays to zero quickly with the increasing $x_{\text{wake}}$. 
For $F_{c1,p}$ and $F_{s1,p}$, the weighted integral of $Q$ (blue line) grows mainly in the range $x_{\text{wake}}\le1.5d$, a region that just includes the vortex V0, and converges to a steady value for $x_{\text{wake}}\ge3d$. For $F_{c2,p}$ and $F_{c3,p}$, their curves remain almost constant for $x_{\text{wake}}\ge0.8d$, and therefore they receive main contributions from the $Q$ structures upstream of $x=0.8d$. $F_{s2,p}$ and $F_{s3,p}$ are small in magnitude and not important to the surface pressure.

The spatial distributions of the volume integrands $-2\rho Q\varphi_{cm}$ and $-2\rho Q\varphi_{sm}$, which represent contributions from local flow structures, are shown in figure~\ref{fig:projQm}, where the vorticity field is also shown by black contour lines. 
Because $\varphi_{c1}$ changes sign across $x=0$, the strain rate in front of the leading edge contributes to drag, i.e. a positive value to $F_{c1,p}$, whereas vortical structures in the wake also contribute to drag. In contrast, the boundary layer on the left cylinder surface and the strain region below vortex V0 contribute to thrust. Similarly, since $\varphi_{s1}$ changes sign across $y=0$, the upstream shear layer and the boundary layers on the left half cylinder show opposite signs in the upper and lower planes, leading to cancellation when evaluating their net contribution to $F_{s1,p}$. On the other hand, both the vortex centre of V0 and the strain rate region below it contribute to lift, which is the main reason for the increase of the volume integral of $-2\rho Q\varphi_{s1}$ in $0.8d<x_{\text{wake}}<1.5d$, as indicated in figure~\ref{fig:projFm}(b). Therefore, the formation of V0 and the strain rate region below it is responsible for the asymmetric pressure distribution on the leeward side of the cylinder surface. The $-2\rho Q\varphi_{c2}$ is mainly distributed with negative values in front of the leading edge and near the highest and lowest points. This is consistent with the fact that the effect of $F_{c2,p}$, which is negative, raises the pressure at the leading edge and reduces $p$ near the highest and lowest points, as illustrated by the Fourier expansion in figure~\ref{fig:Fouriersurfp}(a). In addition, the $-2\rho Q\varphi_{c3}$ is mainly distributed with positive values in front of the leading edge and in the boundary layers on the left-half cylinder. These regions are responsible for the further increased pressure at the leading edge through the additional term $F_{c3,p}$ in the Fourier expansion.

\subsection{Force estimation from a discrete and noisy flow field}
Estimating the fluid force from the flow field provides an approach for non-intrusive force measurements. However, its accuracy and applicability are impeded by the requirement for near-wall velocity derivatives or the pressure field \citep{diaz2022application}. In the weighted projection method~\eqref{eq:weightedproject}, these inconveniences can be circumvented by a specially designed weight function.

Consider a divergence-free vector weight function $\pw$ that satisfies $\pw=\pw_B$ on the body surface $\partial B$ and $\pw=\pzero$ on the outer surface $\Sigma$. Here, $\pw_B$ is a constant vector. By substituting $\pW\cdot\pe_x=\pw$ into equation~\eqref{eq:weightedproject} and using equation~\eqref{eq:fwvis4} for $\pF_{w,\text{vis}}$, the total force exerted on the body is obtained as
\begin{equation}\label{eq:noderiv}
\begin{split}
    \pF\cdot\pw_B &= \int_{V_{f}}{-\rho\pa\cdot\pw \rd V} +\int_{V_f}{\mu \pu\cdot\nabla^2\pw \rd V} +  \oint_{\partial B}{-\mu\pu\cdot[\bnabla\pw+(\bnabla\pw)^T]\cdot\pn\rd S} \\
    & +  \oint_{\Sigma}{-\mu\pu\cdot[\bnabla\pw+(\bnabla\pw)^T]\cdot\pn\rd S}.
\end{split}
\end{equation}
Although the acceleration is involved in the volume integration, equation~\eqref{eq:noderiv} does not contain spatial derivatives of the velocity field, nor does it require the pressure field. As demonstrated in the work of \citet{Lynch_2014} and \citet{Jiang_Lee_Smith_Chen_Linden_2020}, both the fluid velocity and the acceleration fields can be accurately acquired through time-resolved PIV techniques; therefore, equation~\eqref{eq:noderiv} provides a feasible approach for force estimation.

The discretisation of equation~\eqref{eq:noderiv} can be performed by splitting the control volume and its boundary into volume elements $V_i$ and surface elements $\partial B_i$ and $\Sigma_i$, respectively. These are
\begin{equation}
    V_f = \bigcup_i V_i, \quad \partial B = \bigcup_i \partial B_i, \quad \Sigma = \bigcup_i \Sigma_i.
\end{equation}
The following coefficients can be introduced as
\begin{equation}\label{eq:elemntweight}
\begin{split}
    \pf_{V,i} &= \int_{V_i}{-\rho\pw\rd V}, \quad \pgg_{V,i} = \int_{V_i}{\mu\nabla^2\pw\rd V},\quad
    \pgamma_{B,i} = \int_{\partial B_i}{-\mu[\bnabla\pw+(\bnabla\pw)^T]\cdot\pn\rd S}, \\
    \pgamma_{\Sigma,i} &= \int_{\Sigma_i}{-\mu[\bnabla\pw+(\bnabla\pw)^T]\cdot\pn\rd S}.
\end{split}
\end{equation}
It is noted that the coefficients $\pf_{V,i}$, $\pgg_{V,i}$, $\pgamma_{B,i}$ and $\pgamma_{\Sigma,i}$ are fixed once the weight function $\pw$ and domain partition are provided.
Assuming the $V_i$, $\partial B_i$ and $\Sigma_i$ are small enough that the integrands in equation~\eqref{eq:elemntweight} do not change sign in the corresponding integral domains, from the generalised mean-value theorem for integrals, the following holds:
\begin{equation}\label{eq:discretenoderiv}
    \pF\cdot\pw_B = \sum_i \bar{\pa}_i\cdot\pf_{V,i} + \sum_i \bar{\pu}_i\cdot\pgg_{V,i} + \sum_i \tilde{\pu}_i\cdot\pgamma_{B,i}  + \sum_i \tilde{\pu}_i\cdot\pgamma_{\Sigma,i}.
\end{equation}
Here, $\overline{(\;)}_i$ denotes a special value in the volume domain $V_i$ and $\widetilde{(\;)}_i$ denotes a special value on the surface domain $\partial B_i$ or $\Sigma_i$. In applications, these special values can be replaced by values at an arbitrary point in each element, resulting in a first-order numerical quadrature method. Additionally, if the weight function $\pw$, as well as its first- and second-order derivatives, is smooth, the sign-preserving assumption can be satisfied by most elements. Violation of this assumption by a small group of elements leads to a first-order numerical error at worst. Therefore, the weighted summation method~\eqref{eq:discretenoderiv} enables force estimation with first-order spatial accuracy. 

\begin{figure}
    \centering
    \begin{overpic}[width=0.45\linewidth]{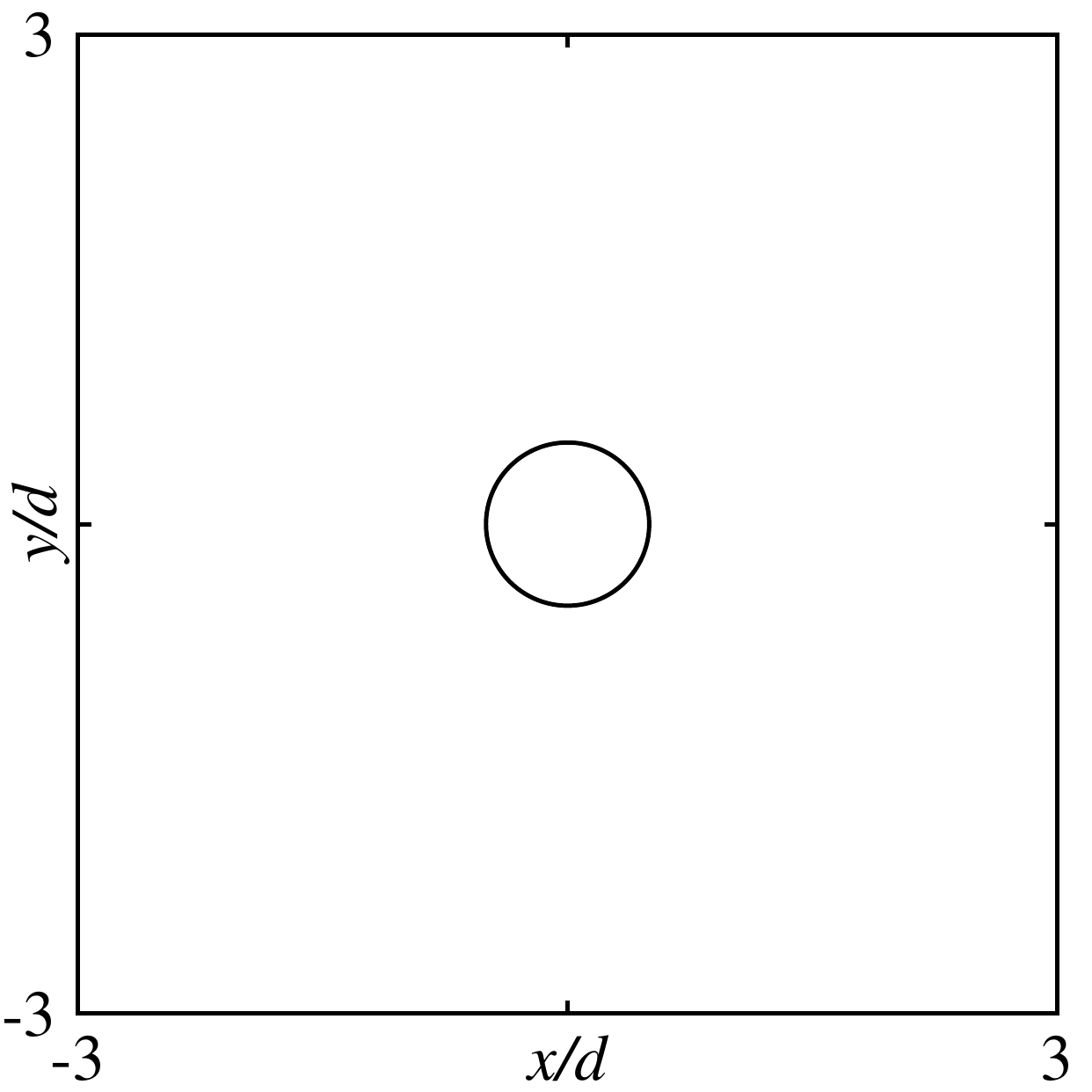}
      \put(-7,97){$(a)$}
      \put(50,38){$\pw=\pw_B$}
      \put(50,90){$\pw=\pzero$}
      \put(20,20){$V_f$}
    \end{overpic}
    \centering
    \begin{overpic}[width=0.45\linewidth]{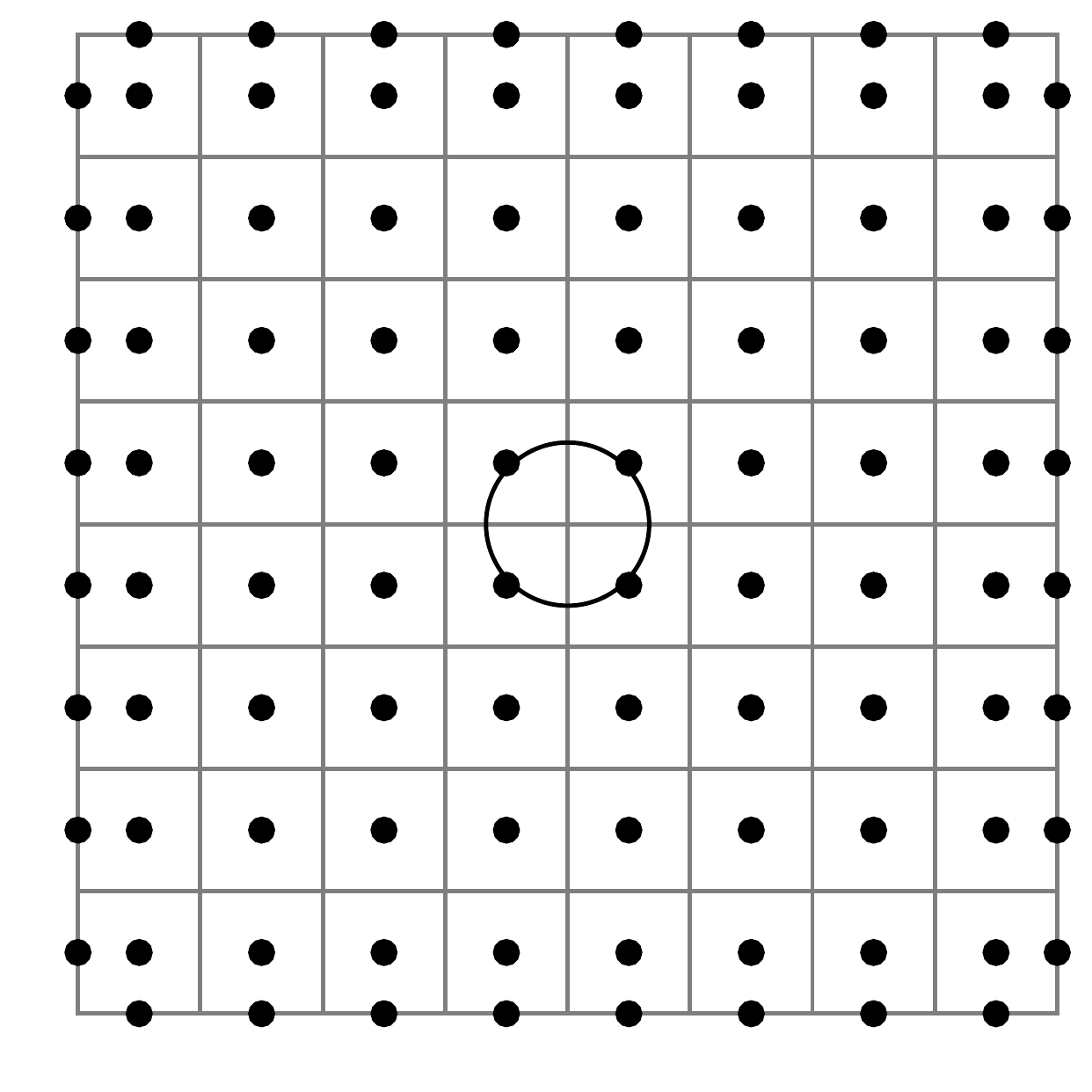}
      \put(-4,97){$(b)$}
    \end{overpic}
\caption{(a) Computational domain and boundary conditions for calculating the vector weight function $\pw$. (b) Grid lines in grey colour show the domain partition, and black dots show the locations of velocity and acceleration sample points. The black circle shows the boundary of the circular cylinder. } \label{fig:Incweightdomain}
\end{figure}

\begin{figure}
\centerline{\includegraphics[width=1\textwidth]{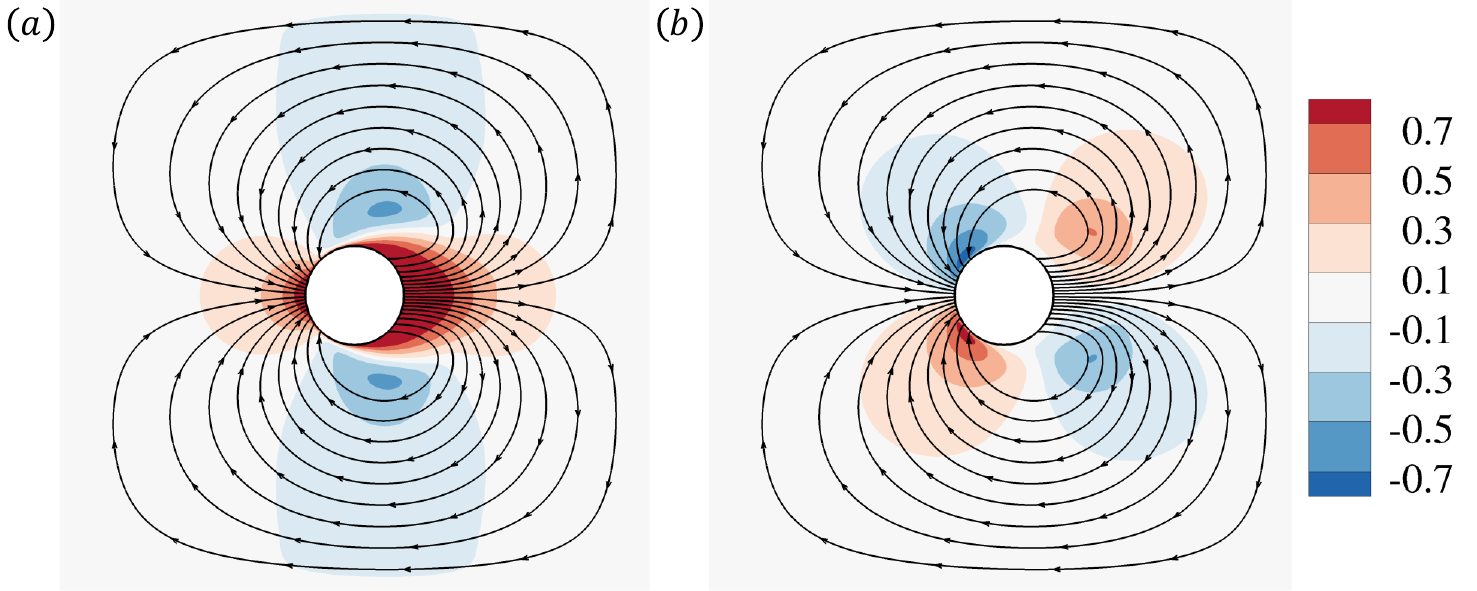}}
	\caption{Colour shows the strength of the vector weight function $\pw$ with $\pw_B=\pe_x$. (a) $\pw\cdot\pe_x$. (b) $\pw\cdot\pe_y$. Black lines with arrows show the streamlines of $\pw$.}
	\label{fig:Incweightfunc}
\end{figure}

Equation~\eqref{eq:discretenoderiv} is tested using the flow past a circular cylinder at $Re=100$. Although the control volume $V_f$ can be set as an arbitrary domain that encloses the cylinder, a medium-sized domain ensures a better spatial resolution in practical applications. Here, $V_f$, which is also the domain of definition of $\pw$, is taken as $-3d\le x\le 3d$, $-3d\le y\le 3d$ and $x^2+y^2\ge 0.25d^2$, see figure~\ref{fig:Incweightdomain}. $V_f$ is divided into $N\times N$ uniform elements. The sample point is placed at the centre of each volume or surface element, as indicated by the black dots. For sample points lying inside the cylinder, their velocity and acceleration are taken as zero. The weight function $\pw$, governed by the divergence-free condition \eqref{eq:continous} and the N-S equations \eqref{eq:momentequation}, is obtained numerically using the incompressible flow solver in Nektar++ \citep{cantwell2015nektar++}, with boundary conditions $\pw=\pzero$ on $\Sigma$ and $\pw=\pw_B$ on $\partial B$, an initial condition $\pw=\pzero$, and a total evolution time of $d/|\pw_B|$. By setting $\pw_B=\pe_x$ or $\pw_B=\pe_y$, the drag and lift can be obtained, respectively. Figure~\ref{fig:Incweightfunc} shows the distribution of $\pw$ with $\pw_B=\pe_x$ in the control volume. Since $\pw$ is generated by a unit-speed blowing on the right half-cylinder and a suction on the left half-cylinder within a time duration of $d/|\pw_B|$, its streamlines reveal two recirculation regions and its magnitude is mainly distributed within the range of $\sqrt{x^2+y^2}\le1.5d$.

\begin{figure}
\centerline{\includegraphics[width=1\textwidth]{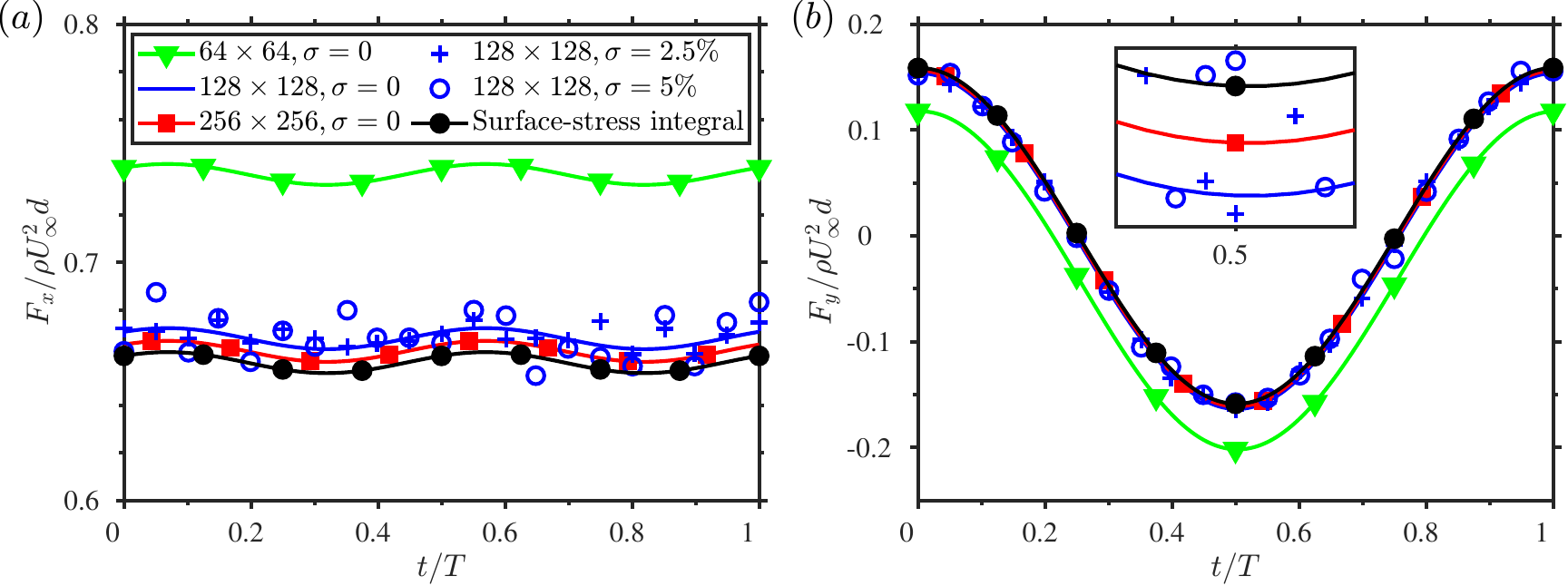}}
	\caption{Dimensionless force calculated by the weighted summation method~\eqref{eq:discretenoderiv} with different spatial resolutions and noise levels. $\sigma$ is the dimensionless standard deviation of the white noise. (a) Drag. (b) Lift.}
	\label{fig:IncweightF}
\end{figure}

\begin{table}
  \begin{center}
  \begin{tabular}{lcc}
  Grid resolution &   Maximum relative error in $F_x$ & Maximum relative error in $F_y$  \\
 \hline
  $64\times 64, \sigma=0$ &   11.6\% & 6.5\% \\
  $128\times 128, \sigma=0$ &  1.5\% & 0.8\%\\
  $128\times 128, \sigma=2.5\%$ & 3.0\% & 2.2\% \\
  $128\times 128, \sigma=5\%$ & 4.3\% & 3.5\% \\
  $256\times 256, \sigma=0$ & 0.7\% & 0.4\%\\
 
  \end{tabular}
  \caption{The maximum relative error of the force calculated using the weighted summation method~\eqref{eq:discretenoderiv}. } 
  \label{tab:discreteerrro}
  \end{center}
\end{table}

To mimic the data measured from experiments, a white noise with a dimensionless standard deviation of $\sigma$ is added to the sampled velocity and acceleration data. 
Forces calculated by the weighted summation method~\eqref{eq:discretenoderiv} with different grid resolutions and $\sigma$ are shown in figure~\ref{fig:IncweightF}. The maximum relative errors are also provided in Table~\ref{tab:discreteerrro}. Here, the relative error is defined based on the maximum force magnitude obtained from the surface-stress integral. At a spatial resolution of $128\times128$, the estimated force shows good agreement with the surface-stress integral. The maximum relative error is 1.5\% for the drag and 0.8\% for the lift. At a higher resolution of $256\times256$, the relative errors in the drag and lift are both halved, in agreement with the first-order spatial accuracy of equation~\eqref{eq:discretenoderiv}. Although the white noise reduces the accuracy of the estimated force, the maximum relative error remains well below 5\% even with a dimensionless standard deviation of $\sigma=5\%$. However, if the spatial resolution is insufficient, such as $64\times64$, significant errors of $11.6\%$ in the drag and $6.5\%$ in the lift are still encountered.

It should be noted that there are still considerable degrees of freedom in the choice of control volume and the weight function $\pw$. The optimisation of the integration domain, element partition and the weight function $\pw$ helps reduce the numerical error. Additionally, a higher-order discretisation of equation~\eqref{eq:noderiv} may also improve the numerical accuracy. These explorations will be addressed in the future.

\section{\label{sec:conclusion}Conclusions}
In this work, a weighted integral framework is established for force diagnostics in incompressible flows. In addition to the weighted surface-stress integral~\eqref{eq:surfFw}, a series of control-volume integral methods are developed, including the general weighted method~\eqref{eq:generalweight}, the weighted projection method with a divergence-free weight function~\eqref{eq:weightedproject}, the weighted projection method with a curl-free weight function~\eqref{eq:potentialprojectvector}, and the three forms of weighted derivative-moment-transformation (DMT) methods, namely equations \eqref{eq:AdvectionDMT}, \eqref{eq:diffusion} and \eqref{eq:boundaryform}. A summary of these methods is presented in Table~\ref{tab:summary}.
The weighted integral methods show benefits in the following two aspects.

First, these methods establish quantitative relationships between the surface stress distribution and the flow field, representing significant progress over current force diagnostics methods, which apply only to the total force and moment. Second, through weighted generalisations of current methods, the physical information captured by each method is clarified and maximised, thereby laying a foundation for a comprehensive analysis of the force mechanisms.

Using the canonical problem of uniform flow past a circular cylinder, the performance of the weighted integral methods is evaluated. In the first application, the kinetic energy balance equation~\eqref{eq:kineticEnergy} for the disturbance flow is obtained by setting the weight function as the disturbance velocity in the weighted projection method with a divergence-free weight function~\eqref{eq:weightedproject}. Results show that the power input by overcoming the cylinder drag is mainly dissipated into heat in the vortex wake, and that a non-negligible portion of kinetic energy is advected far downstream due to the slow decay rate of the disturbance velocity. In the second application, an advection-form weighted DMT method~\eqref{eq:GradPhiAdvectionDMT}, with the weight function being the negative gradient of the auxiliary function introduced by \citet{quartapelle1983force}, is compared with the original advection-form DMT method~\eqref{eq:OriginAdvectionDMT}. The advection-form weighted DMT method recovers the volume force element proposed by \citet{chang1992potential} and also demonstrates superiority in numerical accuracy and convergence speed with the size of the control volume. In the third application, by introducing a sequence of pressure distribution coefficients, a quantitative relationship~\eqref{eq:volFcsmp} between the surface pressure and the $Q$ field is established with the aid of the weighted projection method with a curl-free weight function~\eqref{eq:potentialprojectvector}. Through this relationship, it is found that the strain-rate region in front of the leading edge, the boundary layers, the separated shear layers, and wake vortices within the downstream range of $3d$ are main contributions to the force, while the most recently formed vortex and the strain rate region below it contribute to the lift and the asymmetric pressure distribution about the $x$-axis.
In the last application, a numerical weight function is designed to obtain a force expression involving only the velocity field and the acceleration field, without pressure or spatial derivatives of flow quantities. This force expression~\eqref{eq:noderiv} shows good accuracy and robustness to white noise. 

The weighted integral methods hold promise for applications in a wide range of areas. As shown in the work of \citet{Kangjellyfish2023}, the surface stress distribution has a significant influence on the deformation of the jellyfish-like flexible body, resulting in diverse vortex-shedding patterns and self-propulsive performance. This flow mechanism can be explored using the quantitative integral relationship between the surface pressure and the $Q$ field established in section~\ref{subsec:pdistribution}, with a complete set of basis functions defined on the jellyfish surface. In addition, through optimisations of the control volume, the weight function, and the numerical quadrature method, non-intrusive surface stress measurement techniques of practical value can be developed for experiments. Finally, physical constraints play crucial roles in machine learning methods, as demonstrated in the work of \citet{xie2020prf} and \citet{xie2020pof}. Since the weighted integral methods incorporate information from the Navier-Stokes equations and also impose constraints on the surface pressure and shear stress, they have potential applications in machine learning methods.

\begin{appen}
\section{Nomenclature}
\label{app:nome}
\begin{tabular}{ll}
  \textbf{Symbol} & \textbf{Description} \\
  \hline
  $V_f$ & Finite control volume in the fluid domain  \\
  $V_{f\infty}$ & The whole fluid domain  \\
  $B$ & Volume occupied by the solid body  \\
  $\partial B$ &  Boundary of the solid body $B$, i.e. body surface \\
  $\Sigma$ &  Outer boundary of the control volume \\
  $\partial V_f$ & Boundary of the control volume, $\partial V_f = \partial B + \Sigma$ \\
  $\pn$ & Unit normal vector of $\partial V_f$ pointing outside the control volume $V_f$  \\
  $\hat{\pn}$ & Unit normal vector of $\partial V_f$ pointing inside the control volume $V_f$, $\hat{\pn}=-\pn$ \\
  $\rho$ & Fluid density  \\
  $p$ & Static pressure of the fluid  \\
  $\pu$ & Fluid velocity vector  \\
  $\po$ & Vorticity vector, $\po\coloneq\bnabla\times\pu$  \\
  $\ppl$ & Lamb vector, $\ppl\coloneq\po\times\pu$  \\
  $\pD$ & Strain rate tensor, $\pD\coloneq 0.5\left(\bnabla\pu+(\bnabla\pu)^T\right)$  \\
  $Q$ & The second invariant of the velocity gradient tensor, $Q\coloneq 0.25 |\po|^2-0.5\Vert\pD\Vert^2$  \\
  $\mu$ & Dynamic viscosity  \\
  $\pT$ & Viscous stress tensor, $\pT\coloneq 2\mu\pD$  \\
  $\hat{\ptau}$ & Viscous stress vector on the inner side of $\partial V_f$, $\hat{\ptau}\coloneq \hat{\pn}\cdot\pT$ \\
  $\ptau$ & Viscous stress vector on the outer side of $\partial V_f$, $\ptau\coloneq \pn\cdot\pT$ \\
  $\pa$ & Fluid acceleration \\
  $\pe_i,\pe_x, \pe_y,\pe_z$ & Unit basis vector \\
  $\px$ & Spatial position vector, $\px=x_i\pe_i$ \\
  $k$ & $k+1$ is the space dimension \\
  $\delta_{ij}$ & Kronecker delta \\
  $\epsilon_{ijk}$ & Permutation tensor \\
  $\pW$ & Second-order tensor weight function, $\pW=W_{ij}\pe_i\pe_j$ \\
  $\pw$ & Vector weight function \\
  $w$ & Scalar weight function \\
  $\pF,\; \pM$ & Total force/moment exerted on the body \\
  $\pF_p,\; \pM_p$ & Total pressure force/moment exerted on the body \\
  $\pF_f,\; \pM_f$ & Total friction force/moment exerted on the body \\
  $\ppl_w$ & Weighted Lamb vector, $\ppl_w\coloneq \ppl\cdot\pW$  \\
  $\pF_w$ & Weighted force exerted on the body\\
  $\pF_{w,p}$ & Weighted pressure force exerted on the body\\
  $\pF_{w,f}$ & Weighted friction force exerted on the body\\
  $\pF_{w,\Sigma}$ & Weighted surface-stress integral over $\Sigma$\\
  $\pF_{w,\text{vis}}$ & Volume integral of $-\pT:\bnabla\pW=-T_{ij}(\partial W_{ik}/\partial x_j) \pe_k$ over $V_f$\\
  $\partial_t$ & Eulerian time derivative, $\partial/\partial t$\\
  \hline
\end{tabular}

Subscripts $i,j,k=1,2,3$ and $x, y, z$ indicate the component of vectors and tensors.
\section{Derivative moment transformation}
\label{app_DMT}
The Derivative Moment Transformation (DMT) identities transform the integration of one quantity into the integration of its derivatives \citep{wu2006vorticity}. The following three DMT identities are used in this study.

For the volume integration of a vector function $\pgg$ over $V_f$, there is
\begin{equation}\label{eq:DMTvector}
    \int_{V_f}\pgg\rd V = \frac{1}{k}\int_{V_f}\px\times(\bnabla\times\pgg)\rd V - \frac{1}{k}\oint_{\partial V_f}\px\times(\pn\times\pgg)\rd S.
\end{equation}
Here, $\partial V_f$ is the boundary of the control volume $V_f$, and $k+1$ is the space dimension.

For the integration of a scalar function $\varphi$ over a closed surface $\Sigma$, there is
\begin{equation}\label{eq:DMTscalar}
    \oint_\Sigma{\varphi\pn \rd S} = - \frac{1}{k}\oint_\Sigma{\px\times(\pn\times\bnabla \varphi ) \rd S}.
\end{equation}

In 3-D space, the following equation holds for the integration of a vector function $\pgg$ over a closed surface $\Sigma$,
\begin{equation}\label{eq:DMTcurl}
    \oint_\Sigma{\pn\times\pgg \rd S} = - \oint_\Sigma{\px\times\left[(\pn\times\bnabla)\times\pgg\right] \rd S}.
\end{equation}

\section{Derivation of various forms of $\pF_{w,\mathrm{vis}}$}
\label{app:Fwvis}
It needs to be first clarified that $\pu$ is a divergence-free vector field and that $\pD\coloneq 0.5\left(\bnabla\pu+(\bnabla\pu)^T\right)$ is a symmetric second-order tensor. Equation~\eqref{eq:fwvis1} gives the definition of $\pF_{w,\text{vis}}$,
\begin{equation}\label{eq:appFwvisform1}
    \pF_{w,\text{vis}}\coloneq -2\mu\int_{V_f}{\pD:\bnabla\pW\rd V}.
\end{equation}

To derive equation~\eqref{eq:fwvis2}, the following two vector identities are used,
\begin{equation}\label{eq:appFwvistmp1}
    \bnabla\cdot(\pD\cdot\pW)=(\bnabla\cdot\pD)\cdot\pW+\pD:\bnabla\pW
\end{equation}
and
\begin{equation}\label{eq:appFwvistmp2}
    \bnabla\cdot\pD = 0.5\nabla^2\pu = -0.5 \bnabla\times\po.
\end{equation}
Equation~\eqref{eq:appFwvistmp2} is easily verified since $\bnabla\times(\bnabla\times\pu)=\bnabla(\bnabla\cdot\pu)-\nabla^2\pu=-\nabla^2\pu$. 
Substituting equation~\eqref{eq:appFwvistmp2} into equation~\eqref{eq:appFwvistmp1}, we have
\begin{equation}\label{eq:appFwvistmp3}
    \pD:\bnabla\pW=\bnabla\cdot(\pD\cdot\pW)+0.5(\bnabla\times\po)\cdot\pW,
\end{equation}
and therefore
\begin{equation}\label{eq:appFwvistmp4}
    \pF_{w,\text{vis}} = \int_{V_f}{\left[\bnabla\cdot(-2\mu\pD\cdot\pW)-\mu(\bnabla\times\po)\cdot\pW\right]\rd V}.
\end{equation}
Equation~\eqref{eq:fwvis2} is obtained by applying the divergence theorem to the volume integral of $\bnabla\cdot(-2\mu\pD\cdot\pW)$,
\begin{equation}\label{eq:appFwvistmp5}
    \pF_{w,\text{vis}} = -\mu\int_{V_f}{(\bnabla\times\po)\cdot\pW\rd V}-\oint_{\partial V_f}{\ptau\cdot\pW\rd S}.
\end{equation}

Similarly, from vector identity, $\bnabla\cdot(\po\times\pW)=(\bnabla\times\po)\cdot\pW-\po\cdot(\bnabla\times\pW)$, equation~\eqref{eq:appFwvistmp5} is further transformed into
\begin{equation}\label{eq:appFwvis3tmp1}
    \pF_{w,\text{vis}} = -\mu\int_{V_f}{\left[\bnabla\cdot(\po\times\pW)+\po\cdot(\bnabla\times\pW)\right]\rd V}-\oint_{\partial V_f}{2\mu\pn\cdot\pD\cdot\pW\rd S}.
\end{equation}
After applying the divergence theorem to the volume integral of $\bnabla\cdot(\po\times\pW)$ and using the vector identities $\pn\cdot(\po\times\pW)=(\pn\times\po)\cdot\pW$ and $\pn\times(\bnabla\times\pu)=\bnabla\pu\cdot\pn-\pn\cdot\bnabla\pu$, there is
\begin{equation}\label{eq:appFwvis3tmp2}
    \pF_{w,\text{vis}} = -\mu\int_{V_f}{\po\cdot(\bnabla\times\pW)\rd V}-2\mu\oint_{\partial V_f}{(\bnabla\pu\cdot\pn)\cdot\pW\rd S}.
\end{equation}
Since the following vector identity holds,
\begin{equation}\label{eq:appFwvis3tmp3}
    (\pn\times\bnabla)\cdot(\pu\times\pW) = \epsilon_{ijk}\epsilon_{lmk}n_i\frac{\partial}{\partial x_j}(u_l W_{mr})\pe_r = (\bnabla\pu\cdot\pn)\cdot\pW+(\pn\cdot\pu)(\bnabla\cdot\pW) - \pn\cdot(\pu\cdot\bnabla\pW),
\end{equation}
and the generalised Stokes theorem implies that the integral of $(\pn\times\bnabla)\cdot(\pu\times\pW)$ over a closed surface is zero \citep{wu2006vorticity}, equation~\eqref{eq:fwvis3} is obtained.

With the vector identity $\bnabla\cdot\left[\pu\times(\bnabla\times\pW)\right] = \po\cdot(\bnabla\times\pW)-\pu\cdot\left[\bnabla\times(\bnabla\times\pW)\right]$ and by applying the divergence theorem to the surface integral of $\pn\cdot\left[\pu(\bnabla\cdot\pW)\right]$, equation~\eqref{eq:fwvis3} is transformed into
\begin{eqnarray}
 \nonumber   \pF_{w,\text{vis}} &=& -\mu\int_{V_f}{ \bnabla\cdot\left[\pu\times(\bnabla\times\pW)\right]\rd V} -\mu\int_{V_f}{\pu\cdot\left[\bnabla\times(\bnabla\times\pW)\right]\rd V} \\
  \label{eq:appFwvis4tmp1}  &+& 2\mu \int_{ V_f}{ \bnabla\cdot\left[\pu(\bnabla\cdot\pW)\right] \rd V} - 2\mu \oint_{\partial V_f}{ \pn\cdot(\pu\cdot\bnabla\pW) \rd S}.
\end{eqnarray}
After further applying the divergence theorem to the volume integral of $\bnabla\cdot\left[\pu\times(\bnabla\times\pW)\right]$ and noticing vector identities $\bnabla\cdot\left[\pu(\bnabla\cdot\pW)\right]=\pu\cdot\bnabla(\bnabla\cdot\pW)$ and $\pn\cdot\left[\pu\times(\bnabla\times\pW)\right]=(\pu\pn-\pn\pu):\bnabla\pW$, equation~\eqref{eq:fwvis4} is obtained.

\section{Derivation of the vorticity transport equation from the weighted projection method with a divergence-free weight function}
\label{app:vortransport}
A subspace of the divergence-free weight function that has no net flux on the body surface is considered, i.e.,
\begin{equation}\label{eq:notflux}
    \int_{\partial B}{\hat{\pn}\cdot\pW\rd S} = \pzero.
\end{equation}
From the basic knowledge of calculus, this subspace is equivalent to the functional space $\{\bnabla\times\pPsi\}$ with $\pPsi$ being an arbitrary smooth second-order tensor function. For simplicity, the dot product of the basis vector $\pe_i$ and equation~\eqref{eq:weightedproject} is conducted, resulting in a vector weight function $\bnabla\times\ppsi$ with $\ppsi \coloneq \pPsi\cdot\pe_i$. When adopting equation~\eqref{eq:fwvis2} for the $\pF_{w,\text{vis}}$, the weighted projection method becomes
\begin{equation}\label{eq:projectvector}
\oint_{\partial V_f}{p\pn\cdot(\bnabla\times\ppsi) \rd S} = -\int_{V_{f}}{\rho\pa\cdot(\bnabla\times\ppsi) \rd V} -\mu\int_{V_f}{ (\bnabla\times\po)\cdot(\bnabla\times\ppsi) \rd V}.
\end{equation}
Using vector identities, $\pa\cdot(\bnabla\times\ppsi) = \bnabla\cdot(\ppsi\times\pa)+\ppsi\cdot(\bnabla\times\pa)$ and $(\bnabla\times\po)\cdot(\bnabla\times\ppsi)=\bnabla\cdot[\ppsi\times(\bnabla\times\po)]-\ppsi\cdot(\nabla^2\po)$, the right-hand side of equation~\eqref{eq:projectvector} is transformed using the divergence theorem as
\begin{equation}\label{eq:rhsprojectvector}
-\int_{V_f}{\ppsi\cdot(\rho\bnabla\times\pa-\mu\nabla^2\po)\rd V} + \oint_{\partial V_f}{\ppsi\cdot\left[\pn\times(\rho\pa+\mu\bnabla\times\po)\right]\rd S}.
\end{equation}
Using the vector identity, $p\pn\cdot(\bnabla\times\ppsi)=(\pn\times\bnabla)\cdot(p\ppsi)-(\pn\times\bnabla p)\cdot\ppsi$, the left-hand side of equation~\eqref{eq:projectvector} is transformed via the Stokes theorem as
\begin{equation}\label{eq:lhsprojectvector}
\oint_{\partial V_f}{p\pn\cdot(\bnabla\times\ppsi) \rd S} = -\oint_{\partial V_f}{\ppsi\cdot(\pn\times\bnabla p) \rd S}.
\end{equation}
Combining equations~\eqref{eq:rhsprojectvector} and \eqref{eq:lhsprojectvector}, equation~\eqref{eq:projectvector} is finally transformed to
\begin{equation}\label{eq:projectvector2}
\int_{V_f}{\ppsi\cdot(\rho\bnabla\times\pa-\mu\nabla^2\po)\rd V} - \oint_{\partial V_f}{\ppsi\cdot\left[\pn\times(\rho\pa + \bnabla p + \mu\bnabla\times\po)\right]\rd S} = 0.
\end{equation}
Since $\ppsi$ is an arbitrary vector function, the vorticity transport equation is recovered from the volume integral as
\begin{equation}
\rho\bnabla\times\pa=\rho\frac{\partial\po}{\partial t}+\rho\pu\cdot\bnabla\po-\rho\po\cdot\bnabla\pu = \mu\nabla^2\po.
\end{equation}
The boundary vorticity flux (BVF) that serves as the Neumann boundary condition of the vorticity transport equation is recovered from the surface integral as
\begin{equation}
\mu{\pn}\cdot\bnabla\po = \mu({\pn}\times\bnabla)\times\po +\rho{\pn}\times\pa+{\pn}\times\bnabla p.
\end{equation}

It should be noted that equation~\eqref{eq:notflux} may be violated for local force calculations; given this, the weighted projection method, equation~\eqref{eq:weightedproject}, may contain more information than the vorticity transport equation and the vorticity boundary condition. 

\section{Derivation of the pressure Poisson equation from the weighted projection method with a curl-free weight function}
\label{app:pressurepoisson}
For simplicity, the dot product of the basis vector $\pe_i$ and equation~\eqref{eq:potentialprojectvector} is first conducted, resulting in a vector weight function $\bnabla\phi_i$. The weighted projection method with a curl-free weight function becomes
\begin{equation}\label{eq:Appendprojectphi}
\oint_{\partial V_f}{p\pn\cdot\bnabla\phi_i \rd S} = \int_{V_{f}}{\left(-2\rho Q\phi_i +p\nabla^2\phi_i \right) \rd V} +\oint_{\partial V_f}{\pn\cdot(-\rho\pa+\mu\nabla^2\pu)\phi_i\rd V}.
\end{equation}
Using the vector identity $p\nabla^2\phi_i=\phi_i\nabla^2p + \bnabla\cdot(p\bnabla\phi_i-\phi_i\bnabla p)$, equation~\eqref{eq:Appendprojectphi} is transformed, via the divergence theorem, into
\begin{equation}\label{eq:Appendprojectphi2}
\int_{V_{f}}{\phi_i(\nabla^2p-2\rho Q) \rd V}+\oint_{\partial V_f}{\phi_i\pn\cdot(-\bnabla p - \rho\pa+\mu\nabla^2\pu)\rd S} = 0.
\end{equation}
Since $\phi_i$ is an arbitrary function defined in $V_f$ as well as on $\partial V_f$, the pressure Poisson equation is obtained from the volume integral as
\begin{equation}
\nabla^2 p= 2\rho Q,
\end{equation}
and the Neumann pressure boundary condition is obtained from the surface integral as
\begin{equation}
{\pn}\cdot\bnabla p=-\rho\pa\cdot{\pn}+\mu\nabla^2\pu\cdot{\pn}.
\end{equation}

\section{Derivation of the weighted diffusion- and boundary-form DMT methods}
\label{app:diffu}
To derive the diffusion- and boundary-form weighted DMT methods, an isotropic weight function is required, which is $W_{ij}=w\delta_{ij}$. The volume integrand of equation~\eqref{eq:generalweight} becomes
\begin{equation}\label{eq:deffw2}
    -\rho \pf_w = -\rho w\pa+p\bnabla w-\pT\cdot\bnabla w.
\end{equation}
By applying the DMT identity~\eqref{eq:DMTvector} to the volume integral in equation~\eqref{eq:generalweight}, there is
\begin{equation}\label{eq:tempdiffusion}
 \pF_w = -\frac{\rho}{k}\int_{V_f}{\px\times(\bnabla\times\pf_w) \rd V} + \frac{\rho}{k}\oint_{\partial V_f}{\px\times(\pn\times\pf_w) \rd S} + \oint_\Sigma{(-pw\pn+w\ptau) \rd S}.
\end{equation}
The boundary integrals are denoted by
\begin{equation}\label{eq:FB2}
  \pF_{w,\text{DMT},B} = \frac{\rho}{k}\oint_{\partial B}{\px\times(\pn\times\pf_w) \rd S},
\end{equation}
\begin{equation}\label{eq:FSigma2}
  \pF_{w,\text{DMT},\Sigma} = \frac{\rho}{k}\oint_{\Sigma}{\px\times(\pn\times\pf_w) \rd S}+ \oint_\Sigma{-pw\pn \rd S} + \oint_\Sigma{w\ptau \rd S}.
\end{equation}
Applying the DMT identity~\eqref{eq:DMTscalar} to the surface integral of $-pw\pn$ over $\Sigma$, there is
\begin{equation}\label{eq:FSigma4}
  \pF_{w,\text{DMT},\Sigma} = \frac{1}{k}\oint_{\Sigma}{\px\times\left[\pn\times(\rho\pf_w+\bnabla(pw))\right] \rd S}+ \oint_\Sigma{w\ptau \rd S}.
\end{equation}
Substituting the weighted N-S equations, $\rho\pf_w=-\bnabla(pw)+\bnabla\cdot(w\pT)$, into equations~\eqref{eq:tempdiffusion} and \eqref{eq:FSigma4}, the diffusion-form formula is obtained as follows,
\begin{equation}\label{eq:diffusion2}
 \pF_w = -\frac{1}{k}\int_{V_f}{\px\times\left[\bnabla\times\left(\bnabla\cdot(w\pT)\right)\right] \rd V} + \pF_{w,\text{DMT},B}+\tilde{\pF}_{w,\text{DMT},\Sigma},
\end{equation}
with the surface integral over $\Sigma$,
\begin{equation}\label{eq:FSigma5}
  \tilde{\pF}_{w,\text{DMT},\Sigma} \coloneq \frac{1}{k}\oint_{\Sigma}{\px\times\left[\pn\times(\bnabla\cdot(w\pT))\right] \rd S}+ \oint_\Sigma{w\ptau \rd S}.
\end{equation}

To obtain the boundary-form formula, the DMT identity~\eqref{eq:DMTvector} is first applied with $\pgg=\bnabla\cdot(w\pT)$,
\begin{equation}\label{eq:tmpDMTvec}
     \frac{1}{k}\int_{V_f}\px\times\left[\bnabla\times\left(\bnabla\cdot(w\pT)\right)\right]\rd V - \frac{1}{k}\oint_{\partial V_f}\px\times\left[\pn\times\bnabla\cdot(w\pT)\right]\rd S =\int_{V_f}\bnabla\cdot(w\pT)\rd V= \oint_{\partial V_f}w\ptau\rd S.
\end{equation}
The volume integral in equation~\eqref{eq:diffusion2} is eliminated using equation~\eqref{eq:tmpDMTvec},
\begin{equation}\label{eq:boundary2}
 \pF_w = \frac{1}{k}\oint_{\partial B}{\px\times\left[\pn\times(\rho \pf_w-\bnabla\cdot(w\pT))\right] \rd S}  -\oint_{\partial B}w\ptau\rd S.
\end{equation}
Substituting the weighted N-S equations $\rho\pf_w-\bnabla\cdot(w\pT)=-\bnabla(pw)$ into equation~\eqref{eq:boundary2}, the boundary-form formula is obtained as
\begin{equation}\label{eq:boundary3}
 \pF_w = -\frac{1}{k}\oint_{\partial B}{\px\times\left[\pn\times\bnabla(pw)\right] \rd S}  -\oint_{\partial B}w\ptau\rd S.
\end{equation}
The last term is the weighted friction force $\pF_{w,f}$, since there are $\pn=-\hat{\pn}$ and $\ptau=-\hat{\ptau}$ on the body surface $\partial B$. From the DMT identity~\eqref{eq:DMTscalar}, the first term on the right-hand side is the weighted pressure force
\begin{equation}\label{eq:boundaryFwp}
    - \frac{1}{k}\oint_{\partial B}{\px\times\left[\pn\times\bnabla (pw)\right] \rd S} = \oint_{\partial B}{pw\pn \rd S} = \pF_{w,p}.
\end{equation}
Therefore, the boundary-form force expression \eqref{eq:boundary3} is actually equivalent to the weighted surface-stress integral \eqref{eq:surfFw}.

For the rigid body whose angular velocity is $\pO$, the weighted friction force $\pF_{w,f}$ in equation~\eqref{eq:boundary3} can be further simplified using the generalised Caswell formula, see \citet{wu2006vorticity}. If we denote the relative vorticity as $\po_r\coloneq \po-2\pO$, the viscous stress on $\partial B$ can be expressed as $\hat{\ptau}=\mu\po_r\times\hat{\pn}$, or $\ptau=\mu\po_r\times\pn$. In 3D space, using the DMT identity~\eqref{eq:DMTcurl} with $\pgg=\mu w\po_r$, there is
\begin{equation}\label{eq:FwfMoment3D}
    \pF_{w,f}=-\oint_{\partial B}{\mu w\po_r\times\pn \rd S} = -\mu\oint_{\partial B}{\px\times\left[(\pn\times\bnabla)\times(w\po_r)\right] \rd S}.
\end{equation}
In 2D space, $\po_r=\omega_{rz}\pe_z$. Using the DMT identity~\eqref{eq:DMTscalar}, there is
\begin{equation}\label{eq:FwfMoment2D}
    \pF_{w,f}=-\pe_z\times\oint_{\partial B}{\mu w\omega_{rz}\pn \rd S} = \mu\oint_{\partial B}{\px\left[(\pe_z\times\pn)\cdot\bnabla (w\omega_{rz}) \right] \rd S}.
\end{equation}

\section{Insensitivity of diffusion-form weighted DMT method to internal disturbances}
\label{app:diffuinsensitive}
Assume the vector $\pu''$ and scalar $p''$ are arbitrary smooth functions that equal zero near the body surface $\partial B$. Their gradients $\bnabla\pu''$, $\bnabla\bnabla\pu''$ and $\bnabla p''$ are also zero on the body surface. $(\pu'',p'')$ can be taken as internal disturbances to the flow field. The disturbed flow quantities are denoted with an asterisk, for example, $\pu^*=\pu+\pu''$, $\pT^*=\pT+\mu(\bnabla\pu''+(\bnabla\pu'')^T)$ and $p^*=p+p''$. When the disturbed velocity $\pu^*$ and pressure $p^*$ are substituted into the diffusion-form weighted DMT method~\eqref{eq:diffusion} or \eqref{eq:diffusion2}, the disturbed weighted force is
\begin{equation}\label{eq:diffusionwrongvel}
 \pF_w^* = -\frac{1}{k}\int_{V_f}{\px\times\left[\bnabla\times(\bnabla\cdot(w\pT^*))\right] \rd V} + \pF_{w,\text{DMT}, B}^*+ \tilde{\pF}_{w,\text{DMT},\Sigma}^*.
\end{equation}
According to equation~\eqref{eq:tmpDMTvec}, there is
\begin{equation}\label{eq:2diffusionwrongvel}
 \pF_w^* = - \oint_{\partial B}w\ptau^*\rd S + \frac{1}{k}\oint_{\partial B}{\px\times\left[\pn\times\left(\rho\pf_w^*-\bnabla\cdot(w\pT^*)\right)\right] \rd S}.
\end{equation}
Because $\pu''$, $p''$ and their gradients are all zero on $\partial B$, the integrand in equation~\eqref{eq:2diffusionwrongvel} equals the undisturbed integrand. Therefore, $\pF_w^*=\pF_w$.
Given that $\pu''$ and $p''$ are arbitrary functions and they do not necessarily satisfy the continuity equation or the N-S equations, it is concluded that the diffusion-form weighted DMT method is insensitive to the internal flow disturbances.

\section{Validation of numerical results}\label{app:validationsimu}
The numerical simulation results for the flow past a circular cylinder at $Re=100$ are validated against previous studies. The Strouhal number is defined as $St=d/(U_\infty T)$, where $T$ is the vortex-shedding period. The lift and drag coefficients are defined as $C_L=F_y/(0.5\rho U_\infty^2 d)$ and $C_D=F_x/(0.5\rho U_\infty^2 d)$, respectively. Table~\ref{tab:validationCFD} presents the Strouhal number $St$, the time-averaged drag coefficient $\overline{C_D}$, and the maximum and minimum lift coefficients $C_{L,\max}$ and $C_{L,\min}$. When the polynomial expansion order is reduced from 6 to 4 and the time step is doubled, the relative differences of the force coefficients are less than $0.006\%$, demonstrating excellent convergence with respect to spatial and temporal resolutions. Furthermore, the present results show good agreement with those reported by \citet{LIU199835}, \citet{DING2004727}, and \citet{gao2019passing} in terms of $St$ and the force coefficients.

\begin{table}
  \begin{center}
\def~{\hphantom{0}}
  \begin{tabular}{p{3.5cm}p{2cm}p{2cm}p{2cm}p{2cm}}
  & $St$ & $\overline{C_D}$ & $C_{L,\max}$ & $C_{L,\min}$\\
 \citet{LIU199835} & 0.165 & 1.35 & -0.339 & 0.339 \\
 \citet{DING2004727} & 0.164 & 1.325 & -0.28 & 0.28 \\
 \citet{gao2019passing} & 0.164 & 1.323 & -0.320 & 0.320 \\
 Present, polynomial order 4, time step $0.0005 d/U_\infty$ & 0.16363 & 1.31573 & -0.31740 & 0.31736 \\
 Present, polynomial order 6, time step $0.00024 d/U_\infty$ & 0.16363 & 1.31574 & -0.31739 & 0.31734 \\
  \end{tabular}
  \caption{Validation of the numerical results for the flow past a circular cylinder at $Re=100$.}
  \label{tab:validationCFD}
  \end{center}
\end{table}

\section{Partition of the pressure field and surface pressure integral}
\label{app:pressurepartition}
In the externally unbounded incompressible flow with a constant fluid density, the pressure is governed by the pressure Poisson equation
\begin{equation}
    \label{eq:ppe}
    \nabla^2p=2\rho Q.
\end{equation}
On the body surface $\partial B$, the pressure satisfies a Neumann boundary condition
\begin{equation}
\label{eq:pnbc}
    \hat{\pn}\cdot\bnabla p=-\rho\hat{\pn}\cdot\pa+\mu\hat{\pn}\cdot\nabla^2\pu.
\end{equation}
At the infinitely far field, all disturbances are assumed to decay to zero, and the pressure approaches a constant value, $p_\infty$.

The pressure field can be partitioned as
\begin{equation}
p=p_Q+p_a+p_{\text{vis}}+p_\infty,
\end{equation}
in which $p_Q$, $p_a$ and $p_{\text{vis}}$ represent the pressure fields generated by the volume source $2\rho Q$, the boundary normal acceleration, and the viscous effect on the boundary, respectively. The three partial pressure fields are governed by
\begin{equation}
    \label{eq:ppeparts}
    \nabla^2p_Q=2\rho Q,\quad \nabla^2p_a=0,\quad \nabla^2p_{\text{vis}}=0,
\end{equation}
and they satisfy boundary conditions
\begin{equation}
\label{eq:pnbcparts}
    \hat{\pn}\cdot\bnabla p_Q=0,\quad \hat{\pn}\cdot\bnabla p_a=-\rho\hat{\pn}\cdot\pa,\quad \hat{\pn}\cdot\bnabla p_{\text{vis}}=\mu\hat{\pn}\cdot\nabla^2\pu,
\end{equation}
on the body surface $\partial B$. The total flux of $p_{\text{vis}}$ on the boundary is zero, because the following equation holds
\begin{equation}
    \oint_{\partial B}\mu\hat{\pn}\cdot\nabla^2\pu\rd S=-\mu\oint_{\partial B}(\hat{\pn}\times\bnabla)\cdot(\bnabla\times\pu)\rd S=0,
\end{equation}
which is obtained from the Stokes theorem. Therefore, $p_{\text{vis}}$ decays to zero at infinity.
In 3D space, both $p_Q$ and $p_a$ approach zero at infinity, whereas in 2D space, $p_Q$ and $p_a$ attenuate at the far field only if the total flux of $p_a$ on $\partial B$ and the total volume integral of $Q$ over $V_{f\infty}$ are zero.

The adjoint equation of the pressure Poisson equation, which is self-adjoint, is
\begin{equation}
    \nabla^2\varphi=0.
\end{equation}
On the body surface $\partial B$, $\varphi$ is assumed to satisfy the boundary condition
\begin{equation}
\label{eq:adjointBC}
    \hat{\pn}\cdot\bnabla \varphi=f.
\end{equation}
With the aid of the solution $\varphi$ to the adjoint equation, the weighted surface integrals of partial pressure fields can be obtained without solving equations~\eqref{eq:ppeparts}. There are
\begin{subequations}
\label{eq:surfaceintegral}
 \begin{align}
    \oint_{\partial B} p_Q f \rd S &= \int_{V_{f\infty}}2\rho Q \varphi\rd V, \\
    \oint_{\partial B} p_a f \rd S &= \oint_{\partial B}-\rho\varphi\hat{\pn}\cdot\pa\rd S, \\
    \oint_{\partial B} p_{\text{vis}} f \rd S &= \oint_{\partial B}\mu\varphi\hat{\pn}\cdot\nabla^2\pu\rd S.
\end{align}
\end{subequations}
Here, it is assumed that $p_Q$, $p_a$ and $p_{\text{vis}}$ all decay to zero at infinity and that the volume integration of $2 \rho Q \varphi$ over $V_{f\infty}$ converges.

\begin{figure}
\centerline{\includegraphics[width=1\textwidth]{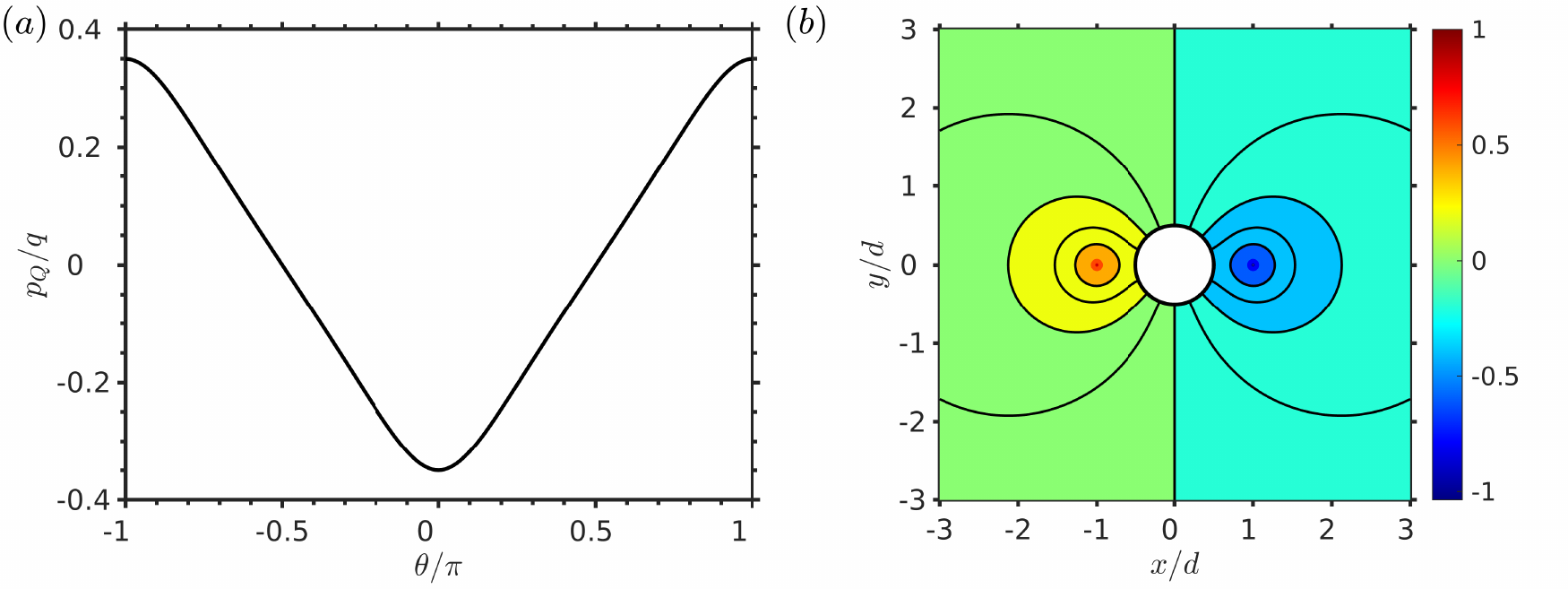}}
	\caption{The pressure field given by equation \eqref{eq:ppointsource} with $\px_0=d\pe_x$ and $\px_1=-d\pe_x$. (a) Pressure distribution on the cylinder surface. $\theta$ is the polar angle. (b) Pressure distribution in the field.}
	\label{fig:Qsuction}
\end{figure}

For example, the pressure field generated by a point source of strength $q\delta(\px-\px_0)$ and a point sink of strength $-q\delta(\px-\px_1)$ outside a circular cylinder with diameter $d$ in 2D space is given by
\begin{equation}\label{eq:ppointsource}
p_Q = \frac{q}{2\pi}\ln|\px-\px_0|+\frac{q}{2\pi}\ln|\px-\px_0'| - \frac{q}{2\pi}\ln|\px-\px_1|-\frac{q}{2\pi}\ln|\px-\px_1'|.
\end{equation}
Here, $\delta(\px)$ is the 2D Dirac delta function, $\px_0'=0.25d^2\px_0/|\px_0|^2$ and $\px_1'=0.25d^2\px_1/|\px_1|^2$. There are $q>0$ and $|\px_0|, |\px_1|>0.5 d$. The integrals of $p_Q$ over the cylinder surface are
\begin{subequations}
\label{eq:pQintegrals}
 \begin{align}
    &\int_{\partial B}{p_Q\rd S} = 0.5dq\ln\left(\frac{|\px_0|}{|\px_1|}\right), \\
    &\int_{\partial B}{-p_Q\hat{\pn}\rd S} = 0.25qd^2\left(\frac{\px_0}{|\px_0|^2}-\frac{\px_1}{|\px_1|^2}\right).
\end{align}
\end{subequations}
Therefore, when the source is distributed closer to the cylinder than the sink, the average surface pressure becomes smaller than the far-field pressure. Moreover, the total pressure force induced by the source manifests as a suction force directed toward the point source, while the sink induces a repulsive force. Figure~\ref{fig:Qsuction} depicts the pressure field for $\px_0=-\px_1=d\pe_x$. Here, the point source in the right half-plane generates a low-pressure region on the adjacent cylinder surface, whereas the point sink in the left half-plane creates a high-pressure region on the proximal surface. Analogously in the flow system, the vortical structure with $Q>0$ produces vortex suction accompanied by a low-pressure surface zone, while a strain-rate region with $Q<0$ generates a repulsive force with a high-pressure area on the nearby body surface.

\end{appen}\clearpage

\begin{bmhead}[Acknowledgments]
An-Kang Gao acknowledges helpful discussions with Jie-Zhi Wu and Zhen Li. We thank the USTC supercomputing center for providing computational resources for this project.
\end{bmhead}

\begin{bmhead}[Funding statement]
The work is supported by the National Natural Science Foundation of China (Grant Nos. 12388101, 12302320, 12293000, and 12293002). This work is also supported by the USTC Startup Program (KY2090000141).
\end{bmhead}

\begin{bmhead}[Declaration of Interests]
The authors report no conflict of interest.
\end{bmhead}

\begin{bmhead}[Data availability statement]
The data that support the findings of this study are available from the corresponding author upon reasonable request.
\end{bmhead}

\begin{bmhead}[Author ORCIDs]
An-Kang Gao https://orcid.org/0000-0002-9805-1388;
Chenyue Xie https://orcid.org/0000-0001-6115-4204;
Xi-Yun Lu   https://orcid.org/0000-0002-0737-6460;
\end{bmhead}

\bibliographystyle{jfm}
\bibliography{jfm}

\end{document}